\documentclass[%
 reprint,
 amsmath,amssymb,
 aps,
]{revtex4-2}

\usepackage{graphicx}
\usepackage{dcolumn}
\usepackage{bm}

\usepackage{amsmath}

\def\refeq#1{{{(\ref{#1})}}}
\newcommand{\out}{{\mathrm{ex}}}
\newcommand{\ins}{{\mathrm{in}}}
\newcommand{\mem}{{\mathrm{m}}}

\newcommand{\kT}{ k_BT}
\newcommand{\sigm}{{\lambda}}

\begin{document}

\preprint{APS/123-QED}

\title{Stationary shapes of axisymmetric vesicles beyond lowest-energy configurations}

\author{Rodrigo B. Reboucas}
\altaffiliation[Present address: ]{Department of Chemical Engineering, University of Illinois at Chicago, Chicago, IL 60608, USA.}
\affiliation{Engineering Sciences and Applied Mathematics, Northwestern University, Evanston, IL 60208, USA.}
\email{rodrigor@uic.edu}
\author{Hammad A. Faizi}%
\altaffiliation[Present address: ]{The Dow Chemical Company, Midland, MI 48611, USA.}
\affiliation{%
 Department of Mechanical Engineering, Northwestern University, Evanston, IL 60208, USA. 
}%

\author{Michael J. Miksis}
\affiliation{Engineering Sciences and Applied Mathematics, Northwestern University, Evanston, IL 60208, USA.}

\author{Petia M. Vlahovska}
\affiliation{Engineering Sciences and Applied Mathematics, Northwestern University, Evanston, IL 60208, USA.}




\date{\today}

\begin{abstract}
We conduct a systematic exploration of the energy landscape of vesicle morphologies within the framework of the Helfrich model. Vesicle shapes are determined by minimizing the elastic energy subject to  constraints of constant area and  volume. The results show that  pressurized vesicles can adopt  higher-energy  spindle-like configurations that require the action of point forces at the poles. If the internal pressure is lower than the external one, multilobed shapes are predicted. We utilize our results to rationalize the experimentally observed spindle shapes of giant vesicles in a uniform AC field.
\end{abstract}

\maketitle

\section{Introduction}\label{sec:introduction} \label{sec: Introduction}
Biomembranes encapsulate cells and cellular organelles and play a key role in regulating essential tasks in natural physiology, such as, efficient transport of oxygen to cells, tissues and organs \cite{herrmann2021extracellular}, signal transmission in neurons \cite{bean2007action}, and immune regulation \cite{van2008membrane,liu2016inflammasome}. These processes are intrinsically complex and rely on a delicate balance between membrane shape transformations and self-generated or externally imposed forces mediated by underlying mechanisms binding the membrane to the cellular cytoskeleton \cite{mulla2018shaping}. In particular, phenomena such as cell division and cell motility may be explained by growth or retraction of cytoskeletal filaments (e.g., actin and microtubules) that generate protrusive forces on the membranes via anchoring proteins \cite{murrell2015forcing,akhmanova2022mechanisms}. Binding-specific proteins allow localized transmission of stresses to the membrane that may lead to microtubule tethering \cite{bezanilla2015cytoskeletal} and formation of spindle-like configurations in a intermediate step of mitotic spindle orientation \cite{di2016regulation}.

\begin{figure}
    \centering
    \includegraphics[scale=.25]{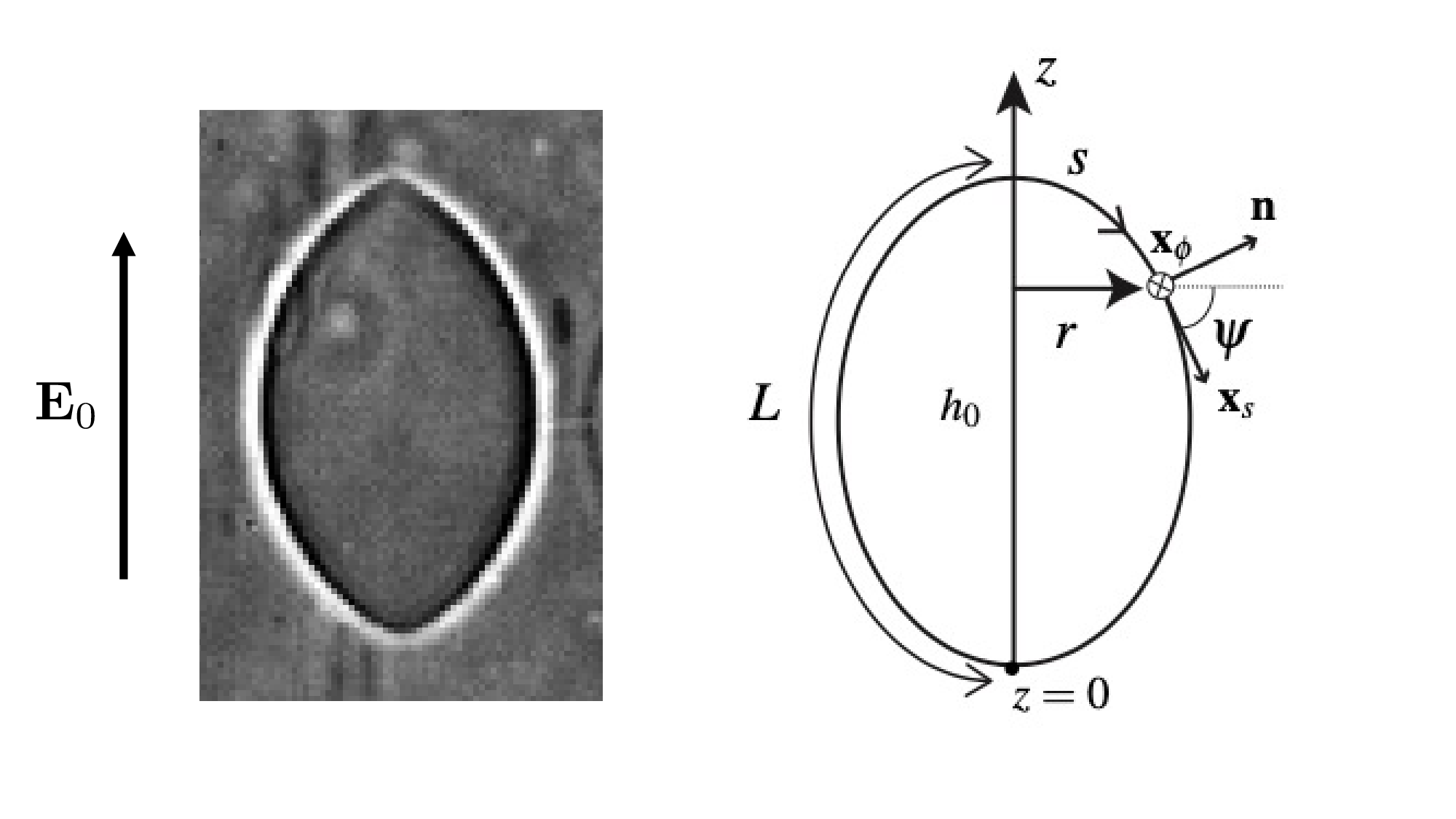}
     \put(-240,120){\large $(a)$}
     \put(-120,120){\large $(b)$}
      \vspace{-.5cm}
    \caption{(a) A giant vesicle in a uniform AC electric field  adopts a spindle shape \cite{Hammad2022}. (b) Schematic showing the coordinate system $(r,\phi,z)$ used to describe the axisymmetric vesicle contour. The shape is parametrized in terms of the tilt angle $\psi$, the arclength $0 \le s \le L$ measured from the north pole, where  the meridional pole-to-pole distance is $L$.  $h_0$ is the length of the vesicle major axis.$\mathbf{x}_{s}$ is the tangent vector along the arclength direction and $\mathbf{x}_{\phi}$ is the azimuthal tangent vector pointing into the page; $\mathbf{n}$ is the outward-pointing normal vector. The origin of the coordinate system is set at the south pole, i.e., at $s=L$.}
    \label{fig:coordinate system}
\end{figure}

The main structural component of biomembranes is a phospholipid bilayer. Giant unilamellar vesicles (GUVs) are cell-sized lipid sacs that self-assemble in aqueous solutions and constitute a popular model to study membrane biophysics \cite{vlahovska2015dynamics}. Pioneering works using vesicles as biomimetic models for living cells were inspired by the discocyte biconcave shape of red blood cells (RBCs) under normal physiological conditions \cite{CANHAM1970,jenkins1977static,ZARDA1977,svetina1989membrane}.  More recently, research has been directed to the understanding of activity-induced vesicle shapes where active particles or filaments confined in GUVs lead to membrane-deformation states varying from tethering to multi-lobed structures where active particles or filaments tend to accumulate in regions of high curvature   \cite{paoluzzi2016shape,vutukuri2020active,peterson2021vesicle}. 
Another class of experimentally observed closed-membrane shapes is spindle-like configurations that may occur, for example, during tether formation induced by the controlled axial growth of confined microtubules \cite{fygenson1997mechanics}, by straining of GUVs embedded in nematic liquid crystals \cite{SaverioPNAS2016,jani2021sculpting}, or, by electrodeformation of vesicles leading to field-induced tubulation at the poles \cite{antonova2016membrane}. Spindle-like shapes, without field-induced tubulation, have been observed recently by modulations of externally applied uniform electric fields \cite{Hammad2022} as shown in Fig.~ \ref{fig:coordinate system}(a). Vesicle shapes induced by localized forces have been identified as higher-energy configurations (compared to unconstrained vesicles) \cite{podgornik1995parametrization,heinrich1999vesicle} and resonate with cell membrane dynamics where activity can be associated with binding of local stresses between the cytoskeleton and the membrane or with asymmetric binding of curvature-inducing proteins \cite{zimmerberg2006proteins}. In the context of electric fields, it has been shown that uniform electric fields acting on cylindrical vesicles of fixed length may drive pearling instabilities; assuming the caps of the cylindrical vesicle to be semi-spherical, an electric tension is coupled to an axial force that orients the vesicle with the field direction and contributes to the global force balance at the interface \cite{sinha2013electric}. We hypothesize that spindle shapes induced by shape distorting electric stresses are analogously driven by modulations of isotropic stresses and point-like forces acting at the poles. 

In this work, we use theory, numerical computation, and experimental analysis to interrogate membrane deformation of GUVs under quasi-steady conditions via modulations of global, isotropic membrane stresses (i.e., membrane tension and transmembrane pressure differences) mimicking the action of external fields. We show a new class of higher-energy vesicle configurations that may have physical implications resembling the dynamic response of biomembranes or synthetic cells driven by self-generated or externally applied forces. More specifically, we concentrate on higher-energy stationary shapes stemming from the classical prolate branch of lowest energy \cite{SeifertBerndlLipowsky1991}, including, but not limited to, spindle-like and tether-like configurations. In order to interpret our theory and numerical results in the context of electric fields, we present an experimental study on the evolution of vesicle shapes in an alternating electric field leading to the formation of spindle-like configurations in high electric field strength regimes (see Fig. \ref{fig:coordinate system}(a)). We propose a numerical methodology that identifies a finite region in a vast parameter space of possible solutions where stationary spindle shapes are identified which is in qualitative agreement with shapes observed experimentally. In \S \ref{sec: modelling membrane mechanics} a description of model assumptions, governing equations, and boundary conditions is presented; section \S \ref{sec:results} contains a collection of vesicle stationary shapes, where results of pressure, tension, length, height, axial force, and bending energy are shown as a function of volume and area; in \S \ref{shapes with positive curvature} we present a numerical mapping of spindle shapes; experimental results are shown in \S \ref{sec:experiments electric field} and concluding remarks are presented in section \S \ref{conclusions}.

\section{Problem formulation}  \label{sec: modelling membrane mechanics}
The thickness of a lipid membrane (e.g., $\approx 5$~nm) is orders of magnitude smaller than the characteristic size of a typical cell or a giant vesicle. Hence, the membrane is treated as a two-dimensional surface embedded in a three-dimensional space \cite{vlahovska2015dynamics}. This separation of length scales allows for a mesoscopic modeling of the membrane where details related to membrane molecular structure are included in material parameters and effective geometric quantities, such as the elastic moduli and spontaneous curvature. 

Typically, the leading order energetic cost of membrane deformation is given by the curvature-elastic energy per unit area \cite{Helfrich1973},
\begin{equation}
    \label{eq bending energy density}
    f=\frac{\kappa}{2}\left(2 H-C_0\right)^2+\kappa_G K
\end{equation}
where $H$ is the mean curvature 
\begin{equation}
    \label{eq mean curvature}
    H=-\frac{1}{2}(c_1+c_2)\,,
\end{equation}
$c_1$ and $c_2$ are the principal radii of curvature, $\kappa$ and $\kappa_G$ are elastic moduli, and asymmetries in the packing of the lipid molecules in the membrane are quantified by the spontaneous curvature, $C_0$. The last term in Eq.~\refeq{eq bending energy density} is the Gaussian curvature,
\begin{equation}
    \label{eq Gaussian curvature}
    K=c_1 c_2\,.
\end{equation}

\subsection{Shape equations}\label{shape equations}
Equilibrium shapes of vesicles have been extensively studied for the past decades and are typically determined by theoretical and numerical approaches that minimize the elastic energy of the membrane \cite{CANHAM1970,lew1972electro,Helfrich1973,evans1974bending,deuling1976curvature,SeifertBerndlLipowsky1991,Seifert1997}. For freely suspended vesicles under constraints of constant area and constant volume, the elastic energy is 
\begin{equation}
    \label{eq Bending Energy no force}
    E'=E_b+ \Sigma \int_{\partial A} \, dA +P \int_{\partial V} dV \,,
\end{equation}
where 
\begin{equation}
    \label{eq Bending Energy}
    E_b=\frac{1}{2}\int_A \kappa (2 H-C_0)^2\,dA\,,
\end{equation}
is the classical Canham-Helfrich bending energy \cite{CANHAM1970,Helfrich1973}, $\Sigma$ and $P=p_{ex}-p_{in}$ are Lagrange multipliers included to enforce the constraints of total area, $A_T$, and total volume, $V_T$, and are often associated with effects of tension within and osmotic pressure difference across the membrane, respectively. Equation~\refeq{eq Bending Energy} is the integral form of the curvature-elastic energy density \refeq{eq bending energy density}. We have neglected the Gaussian curvature energy in \refeq{eq Bending Energy no force} since its integral will only contribute a constant for our problem (e.g., for vesicles having closed surfaces as defined in Ref.~\cite{doCarmo2016differential}). 

The modeling of localized forces leading to membrane protrusions and the experimental realization of vesicle tethers have been an active area of research for many years \cite{hotani1990dynamic,fygenson1997mechanics,bozic1997theoretical,heinrich1999vesicle,PowersPRE2002,derenyi2002formation}. Previous works on axisymmetric vesicle shapes have accounted for the action of an axial force at the poles by including an extra term in the total elastic energy \refeq{eq Bending Energy no force} of the form $-F h_0$, where $F$ is the axial force, and $h_0$ is the height of the vesicle (i.e., the pole-to-pole distance along the axis of symmetry) \cite{bozic1997theoretical,derenyi2002formation}. In an ensemble where the height is fixed, the force enters the energy minimization as a Lagrange multiplier enforcing $h_0$; alternatively, in an ensemble where the force is specified, the height is determined self-consistently \cite{heinrich1999vesicle}. Typically, the length of the vesicle (i.e., arclength pole-to-pole distance) is free to vary and enforces that the total energy of the system is constant. Božič et al. \cite{bozic1997theoretical} and later Heinrich et al. \cite{heinrich1999vesicle} used the generalized area difference model for the elastic energy of membranes and presented results for stationary shapes of prolate freely suspended vesicles deformed axially by a tensile point force yielding prolate-to-spindle shapes with subsequent formation of tethers at the poles. 

Herein, we follow a similar approach where the total elastic energy of closed membranes including the effect of an axial point force, $F$, acting at the north pole is
\begin{equation}
    \label{eq Bending Energy point force}
    G'=E'(H,\Sigma,P)-F \Delta z \vert_{s=0}
\end{equation}
where the south pole is fixed at $s=L$, $\Delta z=(z(0)-z_0)$ is an incremental variation in height at $s=0$ relative to a reference stationary value, $z_0$, measured from the origin of the coordinate system depicted in Fig.~\ref{fig:coordinate system}(b), and  
\begin{equation} 
    \label{eq axial force H}
    F=-4 \pi \kappa (H_s r)\vert_{s=0} 
\end{equation}
is the axial force derived in Appendix \ref{Appendix shape equations}. The axial force \refeq{eq axial force H} is in agreement with a force-and-torque balance derivation given by Eq.~A5 in Ref. \cite{PowersPRE2002}, where transmembrane pressure effects are subleading at the poles. We assume two-fold symmetric shapes such that an equal and opposite point force $-F$ acts at the south pole where $s=L$, and we follow the convention that $F>0$ is a tensile force (i.e., pulling at the poles). Embedded in Eq.~\refeq{eq Bending Energy point force} is the assumption that shape changes occur at a much faster characteristic time scale compared to the rate of application of the constant force, $F$, over an incremental variation in height. 

Minimization of Eq.~\refeq{eq Bending Energy point force} neglecting spatial variations of bending stiffness and asymmetries in the packing of lipid molecules in the membrane (i.e., constant $\kappa$ and $C_0=0$), yields the classical shape equation
\begin{equation}
    \label{Shape Eq normal}
    2\kappa\Delta_{b} H+4\kappa H (H^2-K)-2 H \Sigma -P = 0 \,,
\end{equation}
where $H$ and $K$ are the mean and Gaussian curvatures defined in Eqs.~\refeq{eq mean curvature}-\refeq{eq Gaussian curvature}, respectively, and $\Delta_b$ is the Laplace-Beltrami operator. A derivation  of the shape equation \refeq{Shape Eq normal} using energy minimization is presented in Appendix \ref{Appendix shape equations}, for completeness. The axisymmetric vesicle surface is parameterized by the arclength, $s$, and the principal radii of curvature  
are 
\begin{equation}
    \label{eq meridional curvature}
   c_1=r_s z_{ss}-z_s r_{ss} \,,
\end{equation}
and
\begin{equation}
    \label{eq azimuthal curvature}
   c_2=\frac{z_s}{r} \,,
\end{equation}
where $(r,z)$ are the radial and axial coordinates in cylindrical coordinates illustrated in Fig.~\ref{fig:coordinate system}(b); the subscripts denote differentiation with respect to arclength. Parameterization by arclength introduces an additional local constraint,
\begin{equation}
    \label{eq arclength}
   (r_s)^2 +(z_s)^2 = 1  \,.
\end{equation}

Using Eqs.~\refeq{eq meridional curvature}-\refeq{eq azimuthal curvature} with definitions \refeq{eq mean curvature}-\refeq{eq Gaussian curvature} in the shape equation \refeq{Shape Eq normal} yields a fourth-order partial differential equation in the space variables $(r,z)$ with boundary conditions 
\begin{equation}
    \label{boundary condition r}
    r(0)=0\,, \qquad r(L)=0\,,
\end{equation}

\begin{equation}
\label{boundary condition z}
    z_s(0)=0\,, \qquad z_s(L)=0\,,
\end{equation}
where $L$ is the meridional, arclength pole-to-pole distance of the vesicle. For unconstrained vesicles, the force \refeq{eq axial force H} vanishes at the poles and an additional condition $H_s=0$ is necessary. In Appendix \ref{sec local analysis} we show that when $F=0$ the spatial variables $(r,z)$ are analytic functions of arclength near the poles. If $H_s\neq 0$, then $F$ is finite and a conjugate variable of the vesicle height, $h_0$.

Numerical methods have been used to study the parameter space of stationary vesicle contours as solutions to Eq.~\refeq{Shape Eq normal}. Earlier works used a shooting method where conjugate pairs of variables such as pressure-volume $(P,V)$ and tension-area $(\Sigma,A)$ are adjusted to yield a closed vesicle shape where the length, $L$, and height, $h_0$, of the vesicle are determined self-consistently \cite{peterson1985instability,SeifertBerndlLipowsky1991,BLYTH2004} - see Fig.~\ref{fig:coordinate system}(b) for geometric details. Alternatively, stationary shape equations can be solved implicitly as a two-point boundary value problem in a truncated computational domain with modified boundary conditions \cite{powers2007vesicle}. More sophisticated numerical methods that predict vesicle dynamics in flows also predict equilibrium shapes by a relaxation procedure \cite{veerapaneni2009numerical,salac2011level,laadhari2016adaptive}.  Semi-analytical approaches have also been directed to the modeling of regions of high membrane curvature where elastic boundary layers dominate the dynamics locally \cite{PowersPRE2002,trejo2011effective}. In this work, we use a pseudo-spectral method to solve Eq.~\refeq{Shape Eq normal} numerically as described in section \ref{section: numerical approach} and Appendix \ref{Appendix pseudospectral}. For completeness, in Appendix \ref{sec: Euler-Lagrange discussion}
we show and discuss the connection between the general shape equation \refeq{Shape Eq normal} and the classical system of Euler-Lagrange shape equations widely used in the literature to compute stationary shapes of axisymmetric vesicles (cf.~a comprehensive review by Sefeirt \cite{Seifert1997}). 

\subsection{Dimensionless equations and numerical approach} \label{section: numerical approach}
The dimensionless form of the governing Eqs.~\refeq{Shape Eq normal}, \refeq{eq arclength}, and the force relation \refeq{eq axial force H} are
\begin{equation}
    \label{eq Shape dimensionless normal}
    2 \bar{\Delta}_{b}\bar H+4 \bar{H}(\bar{H}^2-\bar{K})-2 \bar{H} \bar{\Sigma} -\bar{P} = 0\,,
\end{equation}
\begin{equation}
    \label{eq arclength dimensionless}
   (\bar r_s)^2 +(\bar z_s)^2 = 1  \,,
\end{equation}
and 
\begin{equation}
    \label{eq axial force dimensionless}
    \bar F=-\left(\frac{d\bar H}{d\bar s} \bar r \right)_{\bar s=0,\bar L}
\end{equation}
where the over-bars denote dimensionless variables defined by
\begin{eqnarray}
\label{dimensionless lc variables}
&\bar s =\frac{s}{l_c}\,,\quad \bar r=\frac{r}{l_c}\,, \quad \bar z=\frac{z}{l_c}\, ,\quad 
\bar L=\frac{L}{l_c}\,, \nonumber\\ 
& \bar P=\frac{P l_c^3}{\kappa} \, ,\quad
\bar \Sigma=\frac{\Sigma l_c^2}{\kappa}\,, \quad 
\bar H =H\, lc \, ,\quad \bar K=K\,l^2_c \,, 
\end{eqnarray}
and the dimensionless boundary conditions become 
\begin{equation}
    \label{boundary condition r lc}
    \bar r(0)=0\,, \qquad \bar r(\bar L)=0\,,
\end{equation}

\begin{equation}
\label{boundary condition z lc}
    \bar z_s(0)=0\,, \qquad \bar z_s(\bar L)=0\,.
\end{equation}
Accordingly, the dimensionless area, volume, and the axial force are
\begin{equation}
\label{dimensionless A and V lc}
\bar A=\frac{A}{4\pi l_c^2}\, ,\quad \bar V=\frac{V}{(4\pi/3)l^3_c}\,,
\end{equation}

\begin{equation}
\label{dimensionless force lc}
   \bar F =\frac{F\,l_c}{4 \pi \kappa}\,.
\end{equation}

Possible choices for the characteristic length scale are defined in terms of area, volume, pressure, tension, length, axial force, and height. In this work, we use the area-defined characteristic length scale $l_c=\sqrt{A/(4\pi)}$ and equations \refeq{eq Shape dimensionless normal}-\refeq{eq arclength dimensionless} with boundary conditions \refeq{boundary condition r lc}-\refeq{boundary condition z lc} are solved numerically using an implicit pseudo-spectral method \cite{gottlieb1977numerical,canuto2012spectral} where three parameters (e.g., area, volume and length) are specified for constrained vesicles. The resulting non-linear system of algebraic equations is calculated using Newton's method. For unconstrained vesicles, an additional condition $H_s=0$ is enforced that relaxes one of the three parameters specified for constrained vesicles; typically, in this case, the total area and volume are specified, the pressure and tension are determined, and the length and height are calculated self-consistently. Details on the numerical implementation are presented in Appendix \ref{Appendix pseudospectral}. Departure of vesicle shapes from quasi-spherical configurations is usually quantified by the reduced volume \cite{Seifert1997,Vlahovska2019} 
\begin{equation}
    \label{eq reduced volume}
   \nu=\frac{\bar V}{\bar A^{3/2}}\,,
\end{equation}
where $0 <\nu\leq 1$, and physically it represents how deflated the vesicle is according to the ratio of the vesicle true volume to the volume of an equivalent sphere with the same true area. For $l_c=\sqrt{A/(4\pi)}$, the reduced volume equals the dimensionless volume, $\bar V$;  this is the choice of characteristic length scale used in the results presented in section \ref{sec:results}.

\section{Results}\label{sec:results}

In this section, stationary shapes of axisymmetric vesicles are presented. A detailed stability analysis of stationary contours is not pursued in this study and we report vesicle contours with two-fold symmetry only. We present the bending energy of these solutions, and the expectations are that, without any additional external forces, the lowest energy cases are the physically stable. The multidimensional parameter space of the problem spans vesicle shapes with defined surface area, volume, length, height, pressure, tension, axial force, and bending energy. This parameter space is large and our results show a complementary class of stationary shapes stemming from the classical bending energy branch of prolate, unconstrained vesicles that, in certain limits, compare qualitatively with recent experimental results of spindle-like GUV configurations \cite{SaverioPNAS2016,Hammad2022} and with multi-lobed shapes of biological cells \cite{marsh2004direct}.

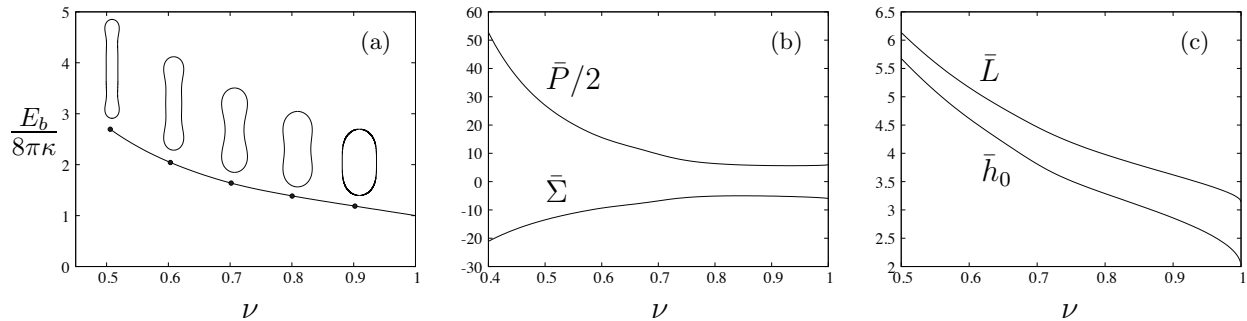
\begin{figure*}
\begin{picture}(250,120)(-15,0)
\put(-135,5){\includegraphics[scale=0.5]{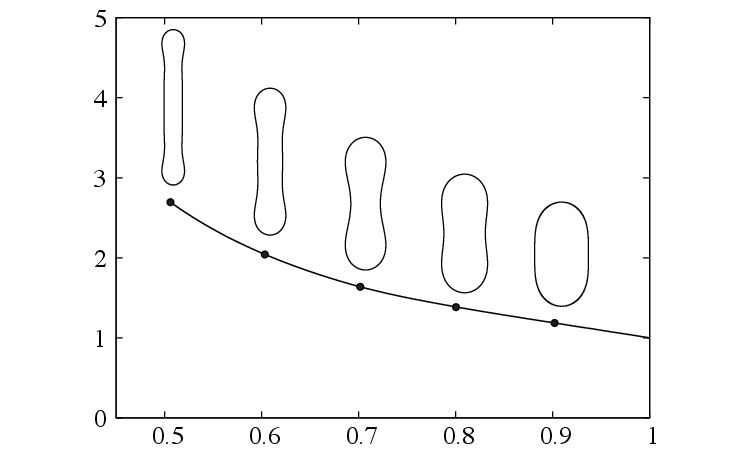} 
\put(-178,55){\Large $\frac{E_b}{8 \pi \kappa}$}
\put(-45,90){(a)}
\put(-90,-12){\large $\nu$}
}
\put(20,5){\includegraphics[scale=0.5]{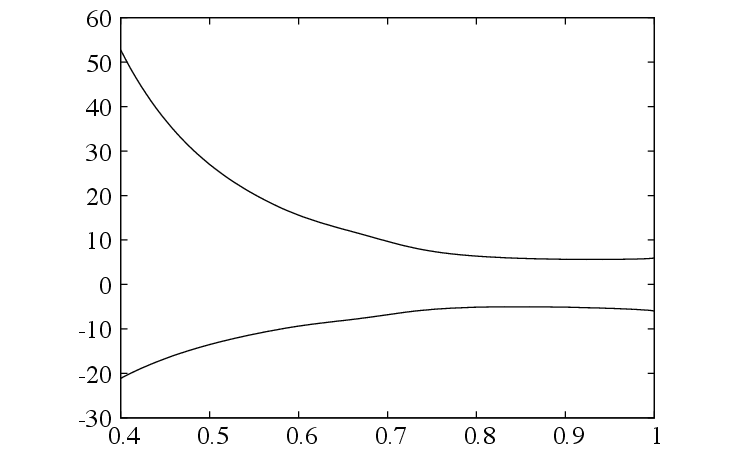} 
\put(-45,90){(b)}
\put(-130,75){\large $\bar P/2$}
\put(-130,32){\large $\bar \Sigma$}
\put(-90,-12){\large $\nu$}
}
\put(175,5){\includegraphics[scale=0.5]{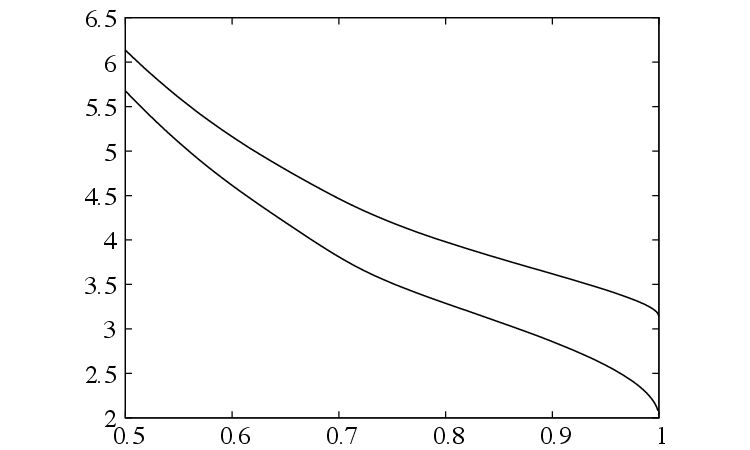}
\put(-121,78){\large $\bar{L}$}
\put(-120,40){\large $\bar{h}_0$}
\put(-45,90){(c)}
\put(-90,-12){\large $\nu$}
}
\end{picture}
\caption{Bending energy of unconstrained vesicles versus reduced volume, $\nu$, (a) (cf. Fig.~8 in Ref.\cite{SeifertBerndlLipowsky1991}); results for pressure and tension (b), where the critical pressure and tension at which spherical shapes become unstable to infinitesimal shape perturbations yielding branches of prolate or oblate vesicles (i.e., $-\bar P/2=\bar \Sigma=6$ at $\nu=1$) is recovered \cite{zhong1987instability}; corresponding values for vesicle length and height (c).}
\label{fig: Seifert results}
\end{figure*}

Contours of prolate, unconstrained vesicles in agreement with classical results of Seifert, Berndl and Lipowsky \cite{SeifertBerndlLipowsky1991} are obtained from the numerical solution of the dimensionless shape equation \refeq{eq Shape dimensionless normal} using the arclength relation \refeq{eq arclength dimensionless} and boundary conditions \refeq{boundary condition r lc}-\refeq{boundary condition z lc} with the additional condition that $H_s=0$ (i.e., $F=0$) at the poles. Values of total area and volume are specified and the Lagrange multipliers $\bar P$ and $\bar \Sigma$, the height $\bar h_0$, and the length $\bar L$ are determined. Figure \ref{fig: Seifert results}(a) shows the bending energy \refeq{eq Bending Energy} of prolate, axisymmetric vesicles as a function of the reduced volume, $\nu$, as defined in \refeq{eq reduced volume}; the corresponding values of pressure, tension, height, and length are shown in Figs.~\ref{fig: Seifert results}(b)-(c). Note that the critical pressure and critical tension at which quasi-spherical vesicles become unstable to infinitesimal shape perturbations (i.e., $-\bar P/2=\bar \Sigma=6$ at $\nu=1$ )\cite{zhong1987instability,SeifertBerndlLipowsky1991} are recovered as shown in Fig.~\ref{fig: Seifert results}(b). 

Constrained vesicle shapes are calculated from solutions to Eqs.~\refeq{eq Shape dimensionless normal}-\refeq{eq arclength dimensionless} and boundary conditions \refeq{boundary condition r lc}-\refeq{boundary condition z lc} where the condition $H_s=0$, enforced for unconstrained vesicles, is relaxed for given surface area and volume; thus, another parameter of the problem needs to be specified. For instance, one could fix the height of the vesicle leading to a non-zero point force $F$ acting at the poles, where the length $L$ of the vesicle is free to vary; or, one could fix the length $L$ of the vesicle and let the height and axial force be determined self-consistently. 

Figure \ref{fig:Energy Fixed A and L, variable V} shows results for the bending energy \refeq{eq Bending Energy} and axial force \refeq{eq axial force H} acting at the poles of axisymmetric vesicles as a function of reduced volume. The solid black line is the prolate branch shown in Fig. \ref{fig: Seifert results}(a) and we investigate vesicle shapes stemming from this curve starting from a vesicle with reduced volume $\nu_0=0.90$ as indicated by point (d) in Fig.~\ref{fig:Energy Fixed A and L, variable V}; the dotted lines emanating from the solid black line represent bending energies of four-, six-, and eight-lobed unconstrained vesicles branching from shape perturbations about a quasi-spherical contour and are added for comparison purposes only. Starting from point (d) in Fig.~\ref{fig:Energy Fixed A and L, variable V}, two branches of solutions arise by fixing the dimensionless area and dimensionless length, and by varying the reduced volume or, equivalently, the dimensionless transmembrane pressure difference. For example, the green line corresponds to pressurized or ``inflated" vesicles where $\bar P$ is decreased (i.e., the internal pressure is increased relative to external one); whereas the red solid line represents ``deflated" vesicles where the external pressure is increased relative to the internal one. Results for the axial force, $F$, are given by the dash-dotted lines in Fig.~\ref{fig:Energy Fixed A and L, variable V} and are calculated using Eq.~\refeq{eq relation eta and force}, where the Lagrange multiplier $\eta$ defined in Appendix \ref{sec: Euler-Lagrange discussion} is determined from relations \refeq{eq Hamiltonian} and \refeq{eq Hamiltonian eq 0} using the numerical values of pressure, tension, the radial coordinate, and the principal curvatures evaluated at the equator (i.e., at $s=L/2$). 

As we increase the internal pressure of the vesicle (or inflate the vesicle) from point (d) in Fig.~\ref{fig:Energy Fixed A and L, variable V}, the bending energy increases followed by a change in sign of the pressure difference across the membrane, where the internal pressure exceeds the external one (i.e., $\bar P<0$), leading to vesicle shapes with high-curvature regions at the poles relative to other parts along the contour \cite{umeda1998theoretical}. This is illustrated in the sequence of shapes (d)-(a) on the right of Fig.~\ref{fig:Energy Fixed A and L, variable V}, where the first shape transition leads to a spindle-like configuration (see Fig.~\ref{fig:Energy Fixed A and L, variable V}(c)) followed by limiting shapes with elongated tips (or tethers) at both poles. Back to the unconstrained prolate shape labeled as (d) in Fig.~\ref{fig:Energy Fixed A and L, variable V}, higher-energy stationary shapes along the red solid line are reported as the internal pressure decreases relative to the external one (or as the vesicle deflates) yielding shapes with increasing number of lobes as seen in the sequence of shapes (d)-(j) shown on the right of the same figure. The points where the dotted black lines for four-, six-, and eight-lobed unconstrained vesicles are tangent to the red solid line correspond to the zeroes of the axial force characterizing transitions in lobe number due to a local change in sign of curvature at the poles according to Eq.~\refeq{eq axial force H}. Initially, the prolate shape dimples at the poles and the point force is compressive. The height of the vesicle decreases and the axial force becomes tensile when $\nu \sim 0.7$ initiating a transition from four- to six-lobed shapes. We hypothesize that this tensile force acting at the poles hinders self-intersection of the contour when the height is close to its minimum value around $\nu\sim 0.5$. 

Moving in the direction of reduced volumes $\nu \lesssim 0.7$, the axial force is tensile and non-monotonic following an increase in vesicle height; a transition from six- to eight-lobed shapes initiates at $\nu \sim 0.77$ when the force becomes compressive again. Similar transitions are expected to happen for even higher modes. The end points in Figs.~\ref{fig:Energy Fixed A and L, variable V} are the final converged shapes obtained from the numerical procedure summarized in Appendix \ref{Appendix pseudospectral}. All shapes marked as (a)-(j) in Fig.~\ref{fig:Energy Fixed A and L, variable V} are shown 
in the Supplementary Material for completeness; relevant parameter values for these shapes are listed in Table \ref{tab: fixed AL variable V}.

\begin{table}
 \caption{Parameter values for length, bending energy, pressure, tension, height, and axial point force for the vesicle shapes marked as (a)-(j) in Fig.~\ref{fig:Energy Fixed A and L, variable V}. Dimensionless variables as defined in Eqs.~\refeq{dimensionless lc variables} and \refeq{dimensionless force lc}.}
 \label{tab: fixed AL variable V}
\begin{ruledtabular}
 \begin{tabular}{ |c|c|c|c|c|c|c| } 
    Shape & $\nu$ & $\bar E_b$ & $\bar P$ & $\bar \Sigma$ & $\bar h_0$ & $2 \bar F$\\
    \hline
    (a) & 0.99  & 3.99 &  -2614 & 1328.5  &  2.72 &  51.44   \\ 
    (b) & 0.97  & 2.02 &  -412.9 &  215.0 &  2.94 &  22.02   \\ 
    (c) & 0.93  & 1.27 &  -42.11  & 23.60 &  3.00 &  5.26   \\ 
    (d) & 0.90  & 1.19 &  11.34  & -5.11  &  2.86 &  0   \\
    (e) & 0.70  & 2.86 &  27.60  & -9.89  &  0.96 &  -1.03   \\
    (f) & 0.56  & 3.37 &  37.60  & -10.29 &  0.17 &  5.20   \\
    (g) & 0.45  & 4.24 &  42.60  & -9.08  &  0.22 &  8.68   \\
    (h) & 0.54  & 5.06 &  51.34  & -12.34 &  1.00 &  6.45   \\
    (i) & 0.85  & 2.74 &  73.34  & -34.03 &  2.60 &  -4.25   \\
    (j) & 0.82  & 4.79 &  98.34  & -39.90 &  1.24 &  1.42   \\
   \end{tabular}
\end{ruledtabular}
\end{table}

The evolution of pressure and tension in response to variations in reduced volume and bending energy are shown in Fig.~\ref{fig:pressure/tension vs nu}. A closer look into the shape evolution near the north pole as one moves up the green solid curve in Fig.~\ref{fig:Energy Fixed A and L, variable V} is depicted in Fig.~\ref{fig: fixed AL variable V, tether}(a), where Figs.~\ref{fig: fixed AL variable V, tether}(b) and (c) show the values of the axial force and vesicle height as a function of reduced volume, respectively. Tether formation occurs after a maximum height is achieved following a monotonic increase of the axial force that grows rapidly for limiting shapes with reduced volume close to one. This divergent behavior of the axial force and of the isotropic effects of pressure and tension shown in Fig.~\ref{fig:pressure/tension vs nu} are a consequence of the geometric limits imposed by fixing the dimensionless area and dimensionless length (i.e., the volume is bounded) and is reflected by the rate of decrease of the tether neck radius according to a scaling between the axial point force and the tether radius $F \sim r_t^{-1}$ in the limit when the tether shape can be approximated by a cylinder with radius $r_t$ \cite{derenyi2002formation}.

This analysis can be directly extended to solution branches starting at different initial reduced volumes, $\nu_0$, along the prolate branch (black solid lines in Figs.~\ref{fig: Seifert results}(a) and \ref{fig:Energy Fixed A and L, variable V}) and the interpretation of the results are qualitatively the same. Note that the classical prolate branch of two-lobed unconstrained vesicles forms an envelope of lowest-energy contours, and spindle-like shapes are observed for dimensionless volumes close to the spherical limit (i.e., $\nu \approx 1$), as illustrated by the contour of Fig.~\ref{fig:Energy Fixed A and L, variable V}(c), in agreement with previous results in the literature \cite{podgornik1995parametrization,heinrich1999vesicle}.

\begin{figure*}
\begin{picture}(250,180)(-15,0)
\put(-148,-45){
\includegraphics[scale=.78]{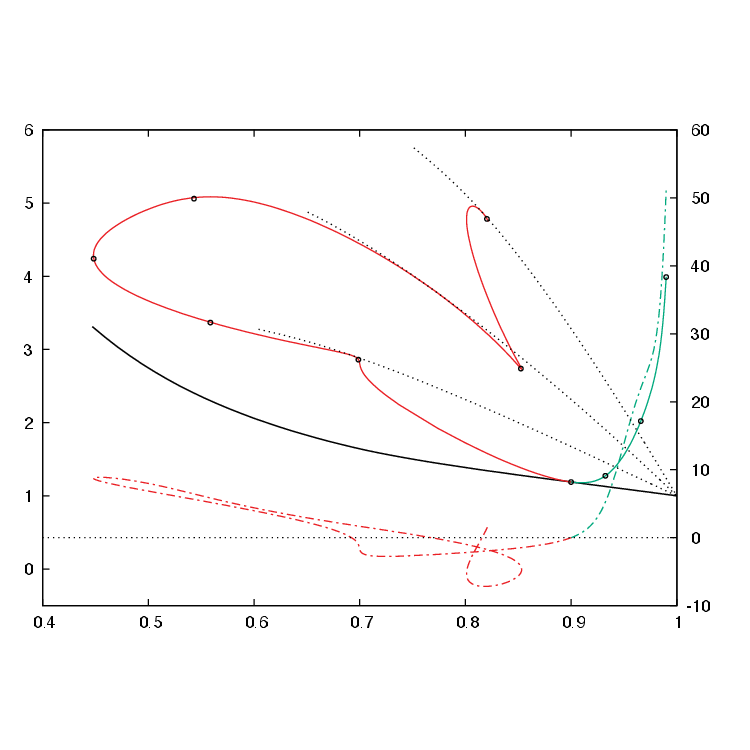}
    \put(-295,145){\Large $\frac{E_b}{8 \pi \kappa}$}
    \put(-140,38){\Large $\nu$}
    \put(-15,140){\Large $\frac{F l_A}{2 \pi \kappa}$}
    \put(-48,178){\small $(a)$}
    \put(-39,118){\small $(b)$}
    \put(-52,89){\small $(c)$}
    \put(-74,90){\small $(d)$}
    \put(-160,139){\small $(e)$}
    \put(-210,150){\small $(f)$}
    \put(-262,182){\small $(g)$}
    \put(-218,213){\small $(h)$}
    \put(-91,133){\small $(i)$}
    \put(-98,205){\small $(j)$}
}
\put(127,102){
\includegraphics[scale=0.2]{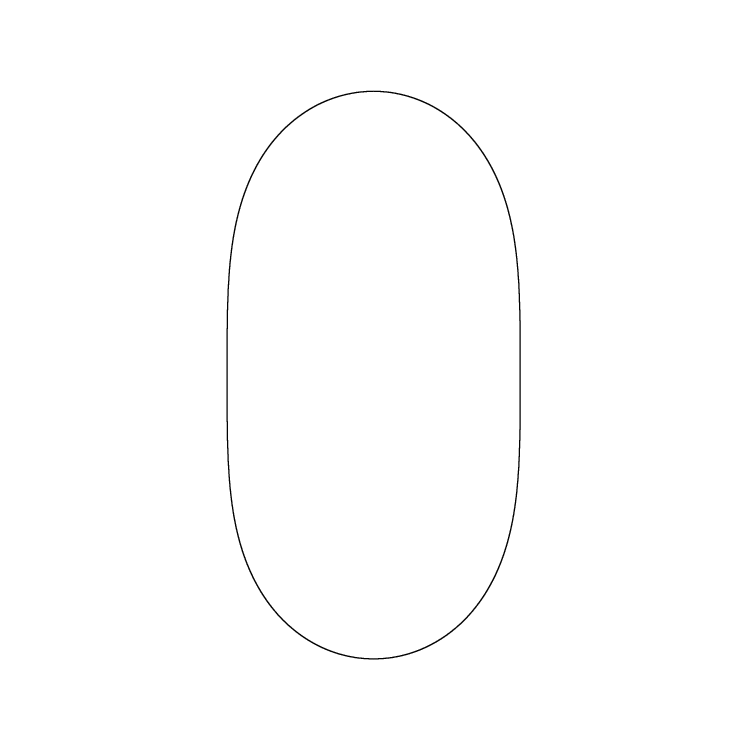} 
\put(-25,68){\small (d)}
\put(-52,-2){\small $\bar P=11.34$}
\put(-52,-14){\small $\bar \Sigma =-5.11$}

}
\put(190,102){
\includegraphics[scale=0.2]{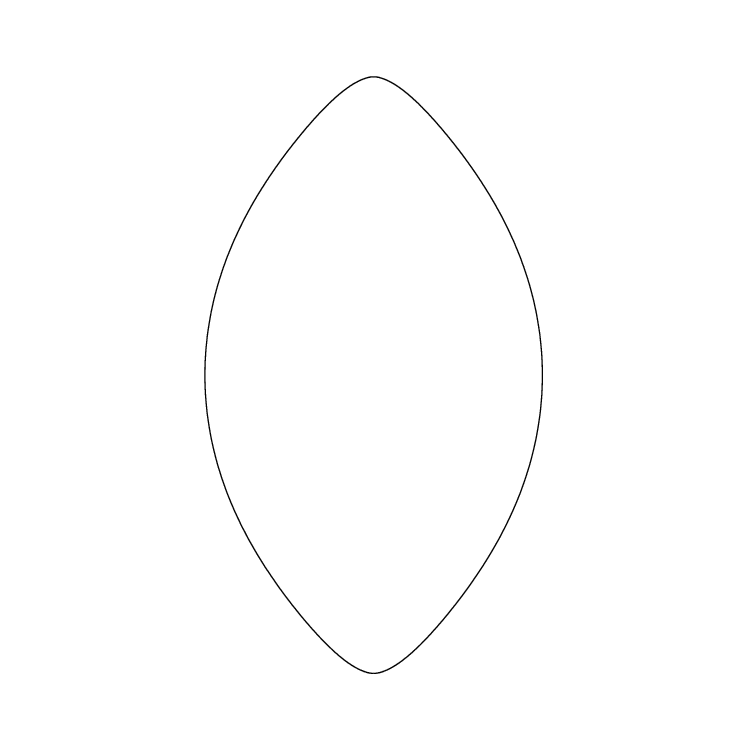} 
\put(-25,68){\small (c)}
\put(-52,-2){\small $\bar P=-42.11$}
\put(-52,-14){\small $\bar \Sigma =23.60$}
}
\put(255,102){
\includegraphics[scale=0.2]{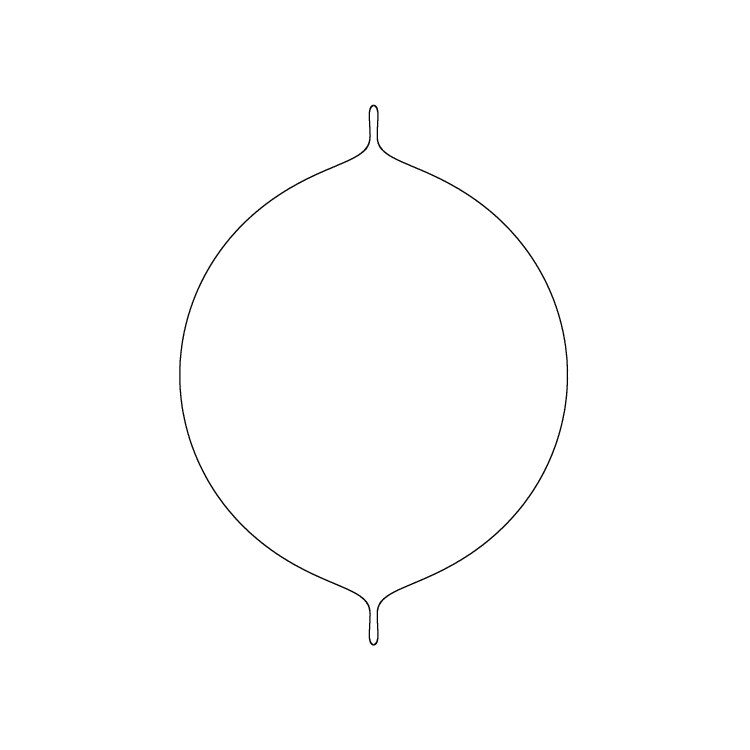} 
\put(-25,68){\small (a)}
\put(-52,-2){\small $\bar P=-2614.0$}
\put(-52,-14){\small $\bar \Sigma =1328.5$}
}
\put(125,2){\includegraphics[scale=0.25]{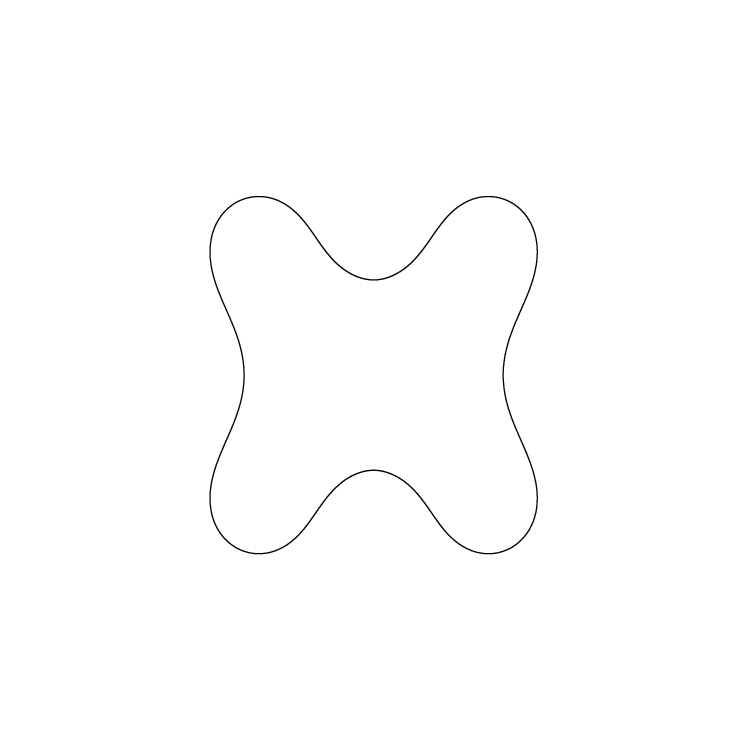}
\put(-25,70){\small (e)}
\put(-66,10){\small $\bar P=27.60$}
\put(-66,-2){\small $\bar \Sigma =-9.89$}
}

\put(187,2){
\includegraphics[scale=0.25]{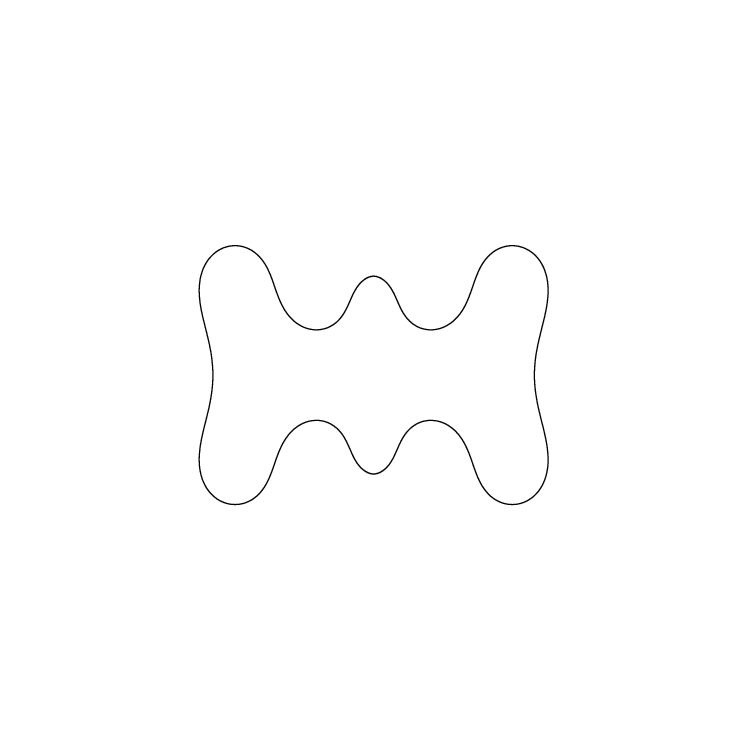}
\put(-25,68){\small (h)}
\put(-66,10){\small $\bar P=51.34$}
\put(-66,-2){\small $\bar \Sigma =-12.34$}
}

\put(255,2){
\includegraphics[scale=0.25]{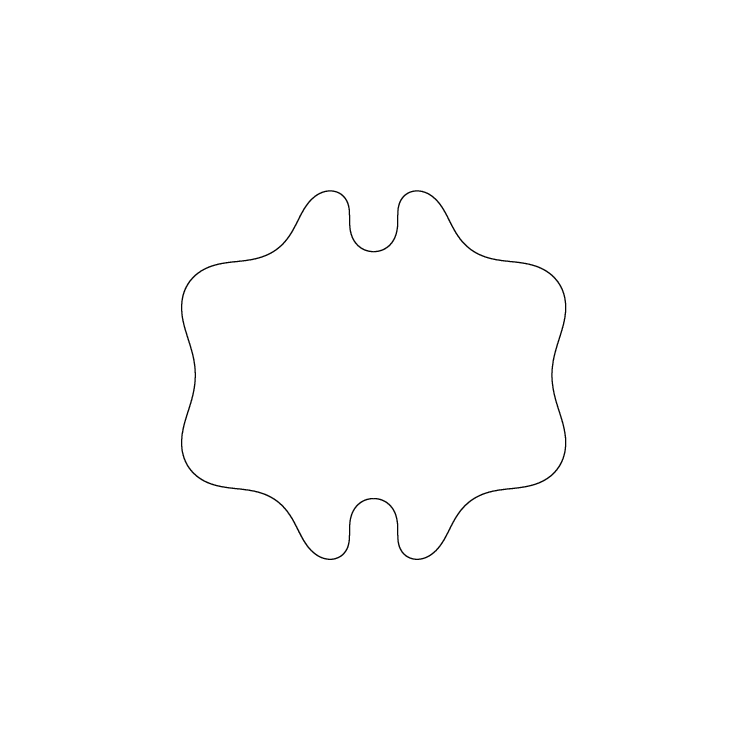}
\put(-25,68){\small (j)}
\put(-66,10){\small $\bar P=98.34$}
\put(-66,-2){\small $\bar \Sigma =-39.90$}
}
\end{picture}
\caption{Bending energy $E_b$ given by Eq.~\refeq{eq Bending Energy} normalized by the bending energy of a unit sphere versus reduced volume, $\nu$; vesicle shapes with fixed area, $\bar A$, and length, $\bar L$, and variable volume.}
\label{fig:Energy Fixed A and L, variable V}
\end{figure*}
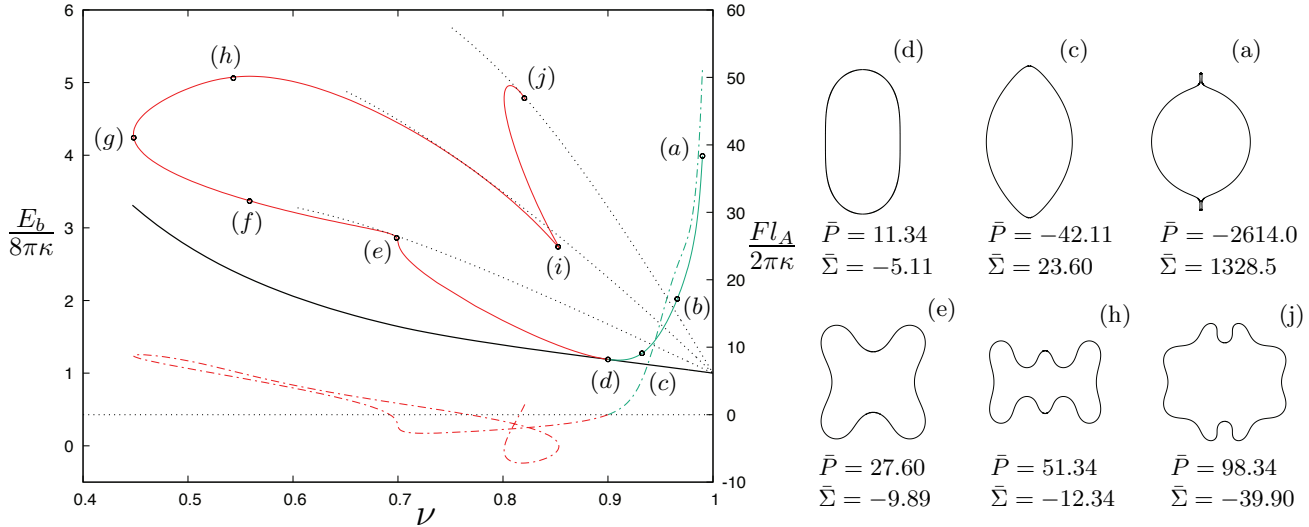

\begin{figure}
\centering
\includegraphics[scale=.68]{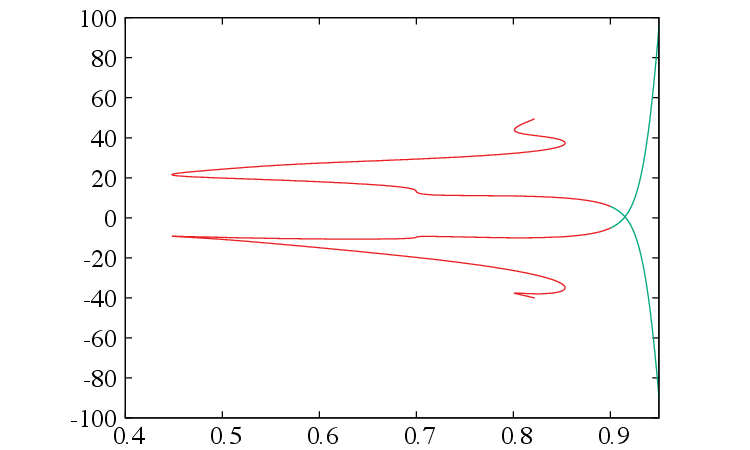} 
\put(-165,105){\Large $\bar P/2$}
\put(-160,48){\Large $\bar \Sigma$}
\put(-115,-10){\Large $\nu$}
\caption{Pressure $\bar P$ and tension $\bar \Sigma$ for vesicle shapes representative of the results shown in Fig.~\ref{fig:Energy Fixed A and L, variable V} shown in part (a); combined effect of pressure and tension (b). Both plots are functions of the reduced volume, $\nu$.}
\label{fig:pressure/tension vs nu}
\end{figure}

\begin{figure*}
\begin{picture}(250,160)(-16,0)
\put(-135,5){\includegraphics[scale=0.5]{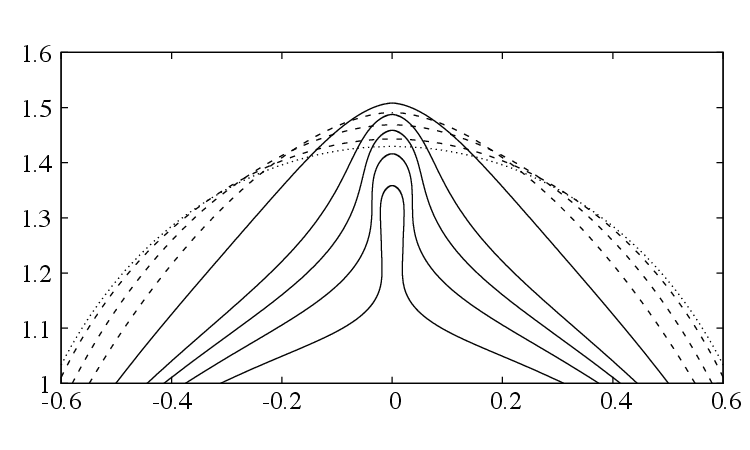} 
\put(-155,80){(a)}
\put(-189,50){\large $\bar z$}
\put(-88,-2){\large $\bar r$}
}
\put(28,5){\includegraphics[scale=0.5]{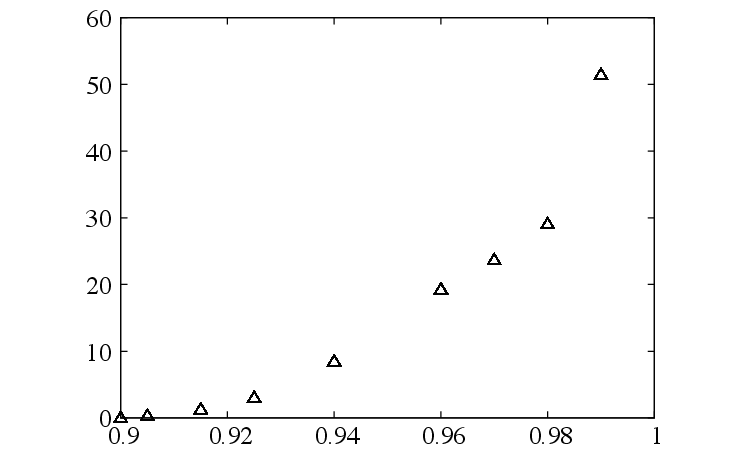} 
\put(-145,90){(b)}
\put(-130,50){\large $F l_A/(2 \pi \kappa)$}
\put(-90,-12){\large $\nu$}
}
\put(175,5){\includegraphics[scale=0.5]{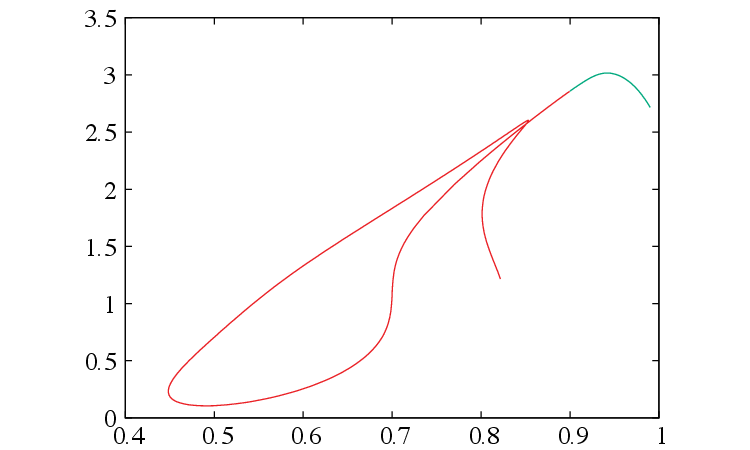}
\put(-120,55){\large $\bar{h}_0$}
\put(-145,90){(c)}
\put(-90,-12){\large $\nu$}
}
\end{picture}
\caption{Shape evolution near the north pole of vesicle shapes along the green curve shown in Fig.~\ref{fig:Energy Fixed A and L, variable V}, (a); corresponding values for the axial force \refeq{eq axial force H} (a) and height $\bar h_0$ (b). All results are plotted versus the reduced volume, $\nu$.}
\label{fig: fixed AL variable V, tether}
\end{figure*}

\section{Numerical mapping of spindle shapes} \label{shapes with positive curvature}
\begin{figure}
    \begin{picture}(200,200)(-56,0)
    \put(-140,2){
    \includegraphics[scale=0.39]{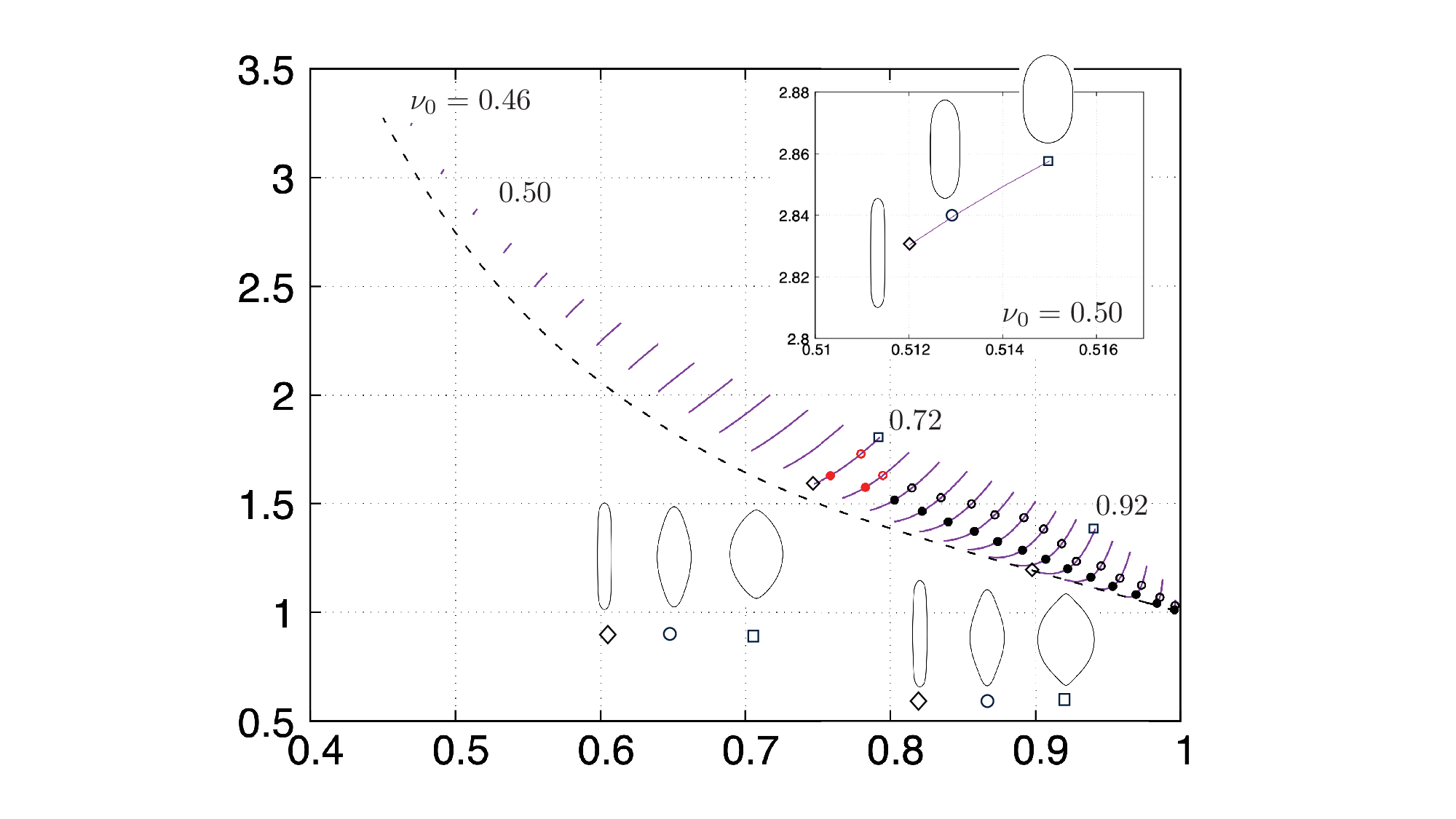}
    \put(-335,112){\huge $\frac{E_b}{8 \pi \kappa}$}
    \put(-190,-2){\Large $\nu$}
    }
    \end{picture}
    \caption{Bending energy, $E_b$, normalized by the bending of a unit sphere versus reduced volume, $\nu$. Bending energy of freely suspended, unconstrained vesicles (dashed line), forced vesicle shapes with positive meridional curvature for all $s$ (continuous lines);  open diamonds and circles indicate extreme shapes along each solid line; filled circles mark the transition to a region of possible spindles and open circles indicate arbitrarily chosen shapes within each branch (spindles for $\nu_0\ge 0.72$). Range of initial reduced volumes of force-free vesicles, $\nu_0$, starting from 0.46 (far-left) to 0.98 (second-last far-right) in 0.2 increments in reduced volume; the last curve on the right starts with $\nu_0=0.995$. Inset shows enlarged region for the $\nu_0=0.50$ energy curve. Vesicle contours along branches starting at $\nu_0=0.50$, $0.72$, and $0.92$ as indicated by the symbols.}
    \label{fig:bending energy positive curvature}
\end{figure}

In this section we propose an approach to map a region of spindle shapes determined numerically within a multidimensional bending energy landscape. We expand and restrict the energy diagram shown in Fig.~\ref{fig:Energy Fixed A and L, variable V} to segments of energy branches of vesicle contours having positive meridional curvature only (i.e., $c_1>0$ for all $0\le s\le L$), which originate from the lowest energy prolate branch of unconstrained vesicles given by the black solid line in Fig.~\ref{fig:Energy Fixed A and L, variable V}. Results are shown in Fig.~\ref{fig:bending energy positive curvature} where the purple lines represent the bending energy of vesicle shapes with positive curvature. The region spans vesicle branches of fixed dimensionless area and dimensionless length (and variable reduced volume) starting from unconstrained vesicles with reduced volumes in the range $0.46\leq \nu_0<1$. Unconstrained vesicles with reduced volumes $\nu_0\gtrsim 0.85$ have positive curvature $c_1$ along the contour and hence the purple lines originate from the black dashed line in Fig.~\ref{fig:bending energy positive curvature} in this range. The first filled circle marks the branch of solutions starting from $\nu_0=0.72$ where the concavitiy of the pressure and tension curves versus reduced volume changes sign. Red filled circles indicate the additional region of possible spindle shapes predicted by numerical analysis relative to the range of spindle shapes observed experimentally marked by black filled circles (for more details on the experimental results, see section \ref{sec:experiments electric field}); open diamonds and squares indicate limiting shapes for a given energy branch, and open circles are arbitrarily chosen shapes within each interval as illustrated by the representative shapes in Fig.~\ref{fig:bending energy positive curvature}.  

\begin{figure}
\centering
\includegraphics[scale=.68]{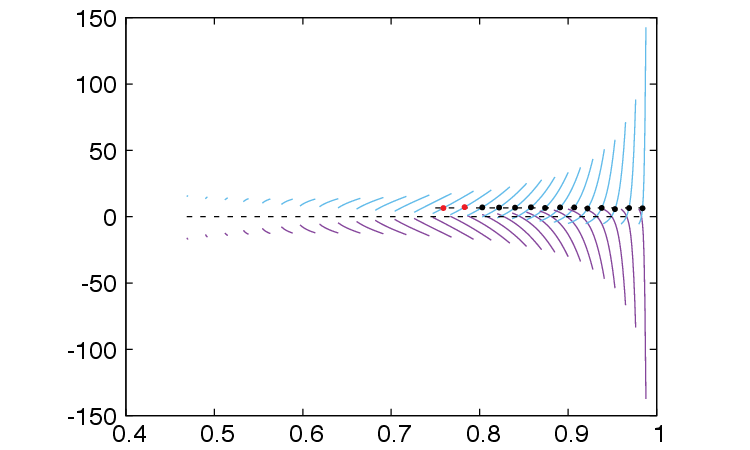} 
\put(-168,50){\Large $\bar P/2$}
\put(-160,92){\Large $\bar \Sigma$}
\put(-180,-12){\Large $\nu$}
\caption{Pressure $\bar P$ and tension $\bar \Sigma$ for vesicle shapes representative of the results shown in Fig.~\ref{fig:Energy Fixed A and L, variable V}; plots are functions of the reduced volume, $\nu$. The dashed line refers to the average tension of all solid circles equal to 6.64.}
\label{fig:pressure tension positive c1}
\end{figure}

Figure \ref{fig:pressure tension positive c1} highlights the aforemenetioned change in concavity of the pressure and tension curves for the energy branch originated from $\nu_0\sim 0.72$ along which $\partial^2 \bar \Sigma/\partial \nu^2|_{\bar A,\bar L}>0$ and $\partial^2 \bar P/\partial \nu^2|_{\bar A,\bar L}<0$. We hypothesize that this inflection point in the tension and pressure curves with respect to reduced volume delineates a region of vesicle shapes that could potentially turn into a spindle configuration as the vesicle is pressurized. Inspection of curvature profiles versus arclength for each shape along the energy branches with $c_1>0$ and $\partial^2 \bar \Sigma/\partial \nu^2|_{\bar A,\bar L}>0$ shows that spindle shapes occur beyond or at points where the concavity of the merdional curvature $c_1$ changes sign at the equator (i.e., at $s=L/2$). All the filled circles in Figs.~\ref{fig:bending energy positive curvature} and \ref{fig:pressure tension positive c1} mark this change in concavity of $c_1$. 

\begin{figure}
    \begin{picture}(200,180)(-68,2)
    \put(-120,2){
    \includegraphics[scale=0.3]{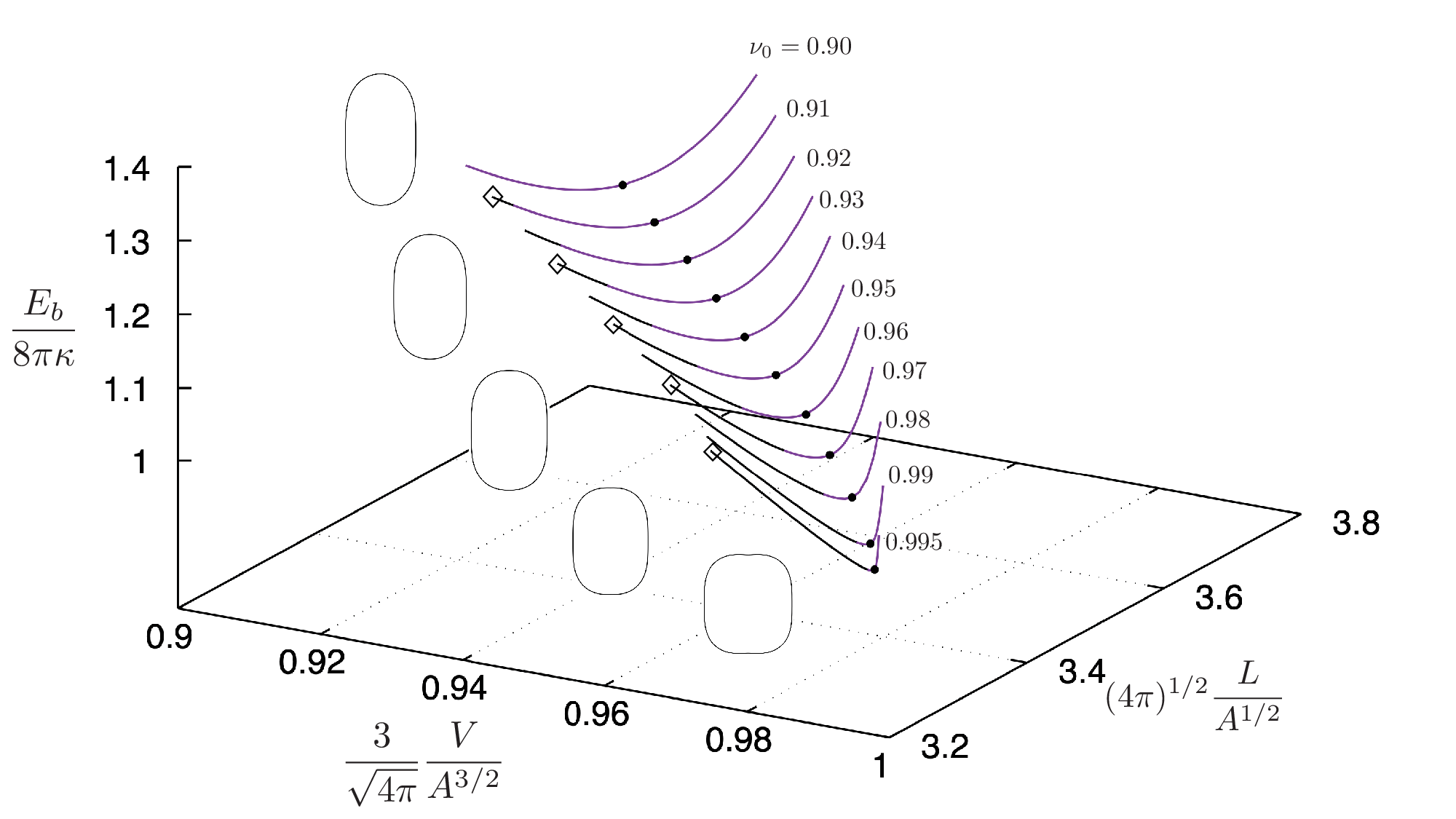}
    }
    \end{picture}
     \caption{Three-dimensional version of Fig.~\ref{fig:bending energy positive curvature} including the variation of length the vesicles as indicated. Black line represent shapes of vesicles with dimpled regions at the poles; extreme contours marked by open diamonds are shown on the right, where $\nu_0$ is the reduced volume of the initial freely suspended vesicle of each solution branch.}
    \label{fig:3D energy curvature}
\end{figure}

Figure \ref{fig:3D energy curvature} shows a three-dimensional version of the bending energy diagram depicted in Fig.~\ref{fig:bending energy positive curvature} for reduced volumes, $\nu\ge0.90$, where the extra dimension is the vesicle length. The additional black solid lines in Fig.~\ref{fig:3D energy curvature} represent vesicle shapes with dimpled regions at the poles only (i.e., where $c_1\le 0$ locally) and positive meridional curvature everywhere else. Dimpled shapes are obtained by initially compressing unconstrained vesicles (or by deflating them via modulations of pressure and tension as shown by the red solid curve in Fig.~\ref{fig:Energy Fixed A and L, variable V}). Representative shapes are shown in Fig.~\ref{fig:3D energy curvature} and are marked by open diamonds on the plot as a reference. Figure \ref{fig:3D energy curvature} indicates that spindle and dimpled shapes at the poles may coexist in a finite region of the bending energy landscape due to perturbations in axial forces or isotropic stresses about unconstrained vesicle configurations.

Our numerical results show that spindle-like shapes are characterized by configurations where the internal pressure is greater than the external pressure (i.e., $\bar P<0$) \cite{umeda1998theoretical}, whereas results for unconstrained vesicles present excess external pressure, as seen in Fig.~\ref{fig: Seifert results}(b). This indicates that by controlling the membrane pressure-tension response to external stimuli, one could design an experimental system where higher-energy spindle shapes could be observed. In fact, spindle shapes reported in Fig.~\ref{fig:Energy Fixed A and L, variable V} are in qualitative agreement with spindle configurations of GUVs observed experimentally and driven by different mechanisms, e.g., by the axial growth of microtubules \cite{fygenson1997mechanics}, the strain of GUVs in nematic liquid crystals \cite{SaverioPNAS2016,jani2021sculpting}, and, more recently, by modulations of an externally applied, uniform electric field \cite{Hammad2022} as discussed in Section \ref{sec:experiments electric field}.

\section{Experimental observation of a sphere-to-spindle transition of GUVs in a uniform electric field}\label{sec:experiments electric field} 

GUVs were prepared from DOPC (dioleoylphosphatidylcholine) using the standard procedure of electroformation \cite{faizi_2022_Langmuir}. GUVs response to a uniform AC electric field was studied in a custom-made chamber \cite{faizi2021electromechanical}. In an AC uniform electric field, vesicles display a frequency-dependent ellipsoidal deformation \cite{Aranda:2008,Vlahovska-Dimova:2009,salipante2012electrodeformation,Aleksanyan_2023_PhysicsX}. A vesicle adopts a prolate spheroidal shape at low frequencies. If the vesicle is filled with a solution that is less conducting than the suspending medium, i.e., the ratio of inner and outer conductivities $\Lambda=\sigm_\ins/\sigm_\out<1$, increasing the frequency induces an oblate spheroidal deformation. The prolate-to-oblate transition occurs a critical 
 frequency \cite{yamamoto2010stability}
\begin{equation}
\label{eq critical fc}
f_{c}=\frac{\sigm_\ins}{2\pi{a}\,C_\mem } \left[\left(1-\Lambda\right)\left(\Lambda+3\right)\right]^{-{1}/{2}}\,,
\end{equation}
where $a$ is the initial radius of the vesicle, and $C_m$ is the membrane capacitance; typically, the critical frequency is in the range of 10 to 100 kHz. At the critical frequency, the vesicle is spherical at low field strengths. However, increasing the field amplitude induces several responses, including transition to a spindle-like shape.

\subsection{Vesicle responses to an electric field with an increasing amplitude} \label{sec:comparison with exp}
\begin{figure}
    \begin{picture}(190,190)(-68,-2)
    \put(-100,-5){
    \includegraphics[scale=0.35]{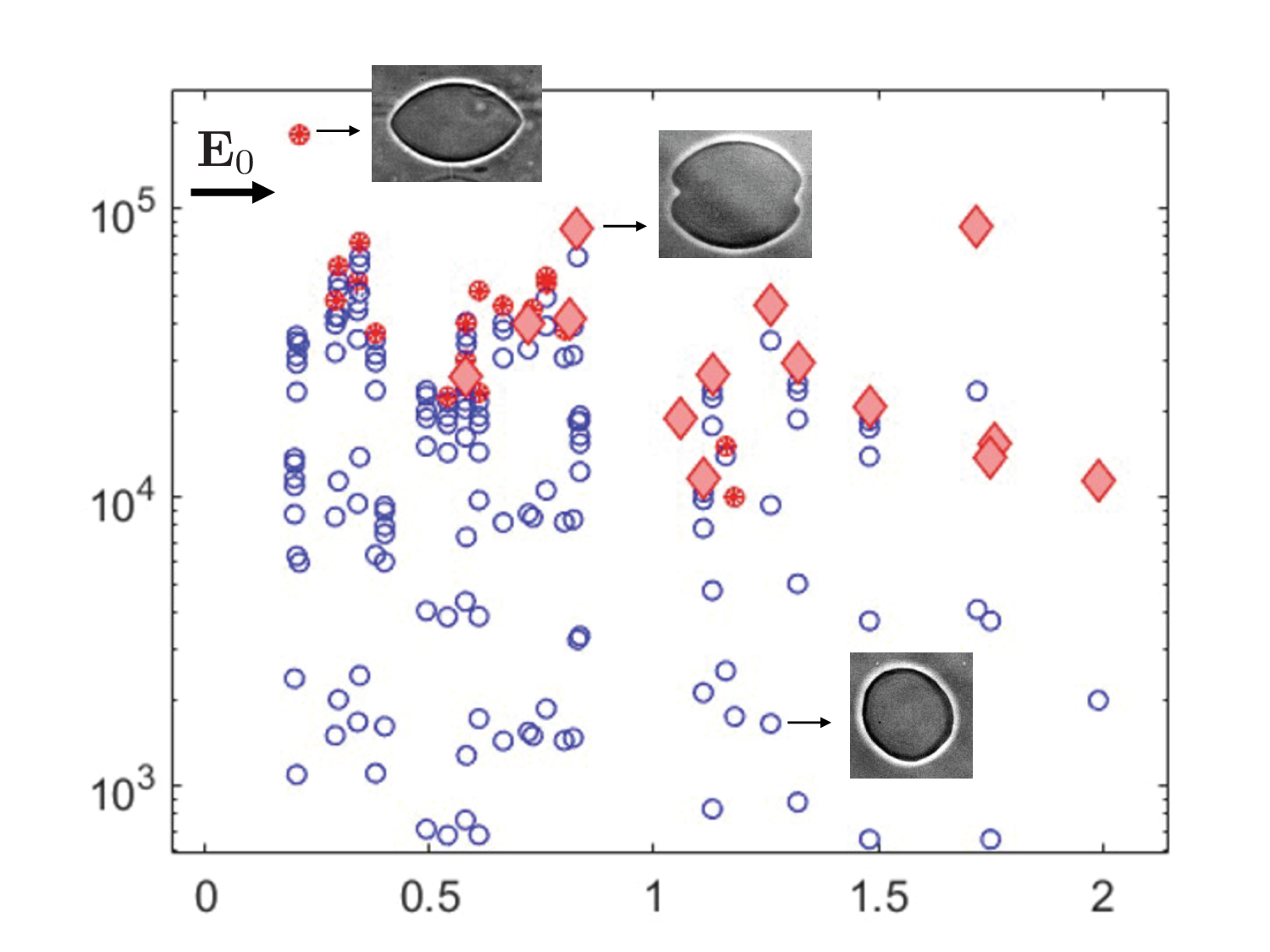}
    \put(-250,110){\Large $Ca_{el}$}
    \put(-135,-2){\Large $\Delta$}
    }
    \end{picture}
    \caption{Phase diagram of vesicles at the critical frequency \refeq{eq critical fc} in AC electric field as a function of excess area $\Delta$ and electrical capillary number $Ca_{el}$. The inner and outer solutions are  1 mM NaCl (conductivity $126\pm1\,\mu{S}\,{cm^{-1}}$) and 1.5 mM NaCl (conductivity $186\pm1\,\mu{S}\,{cm^{-1}}$), respectively. Open blue circles represent stable vesicles where the contours fluctuate about their mean quasi-spherical shape. The filled red circles represent the spindle-like vesicles, and red diamonds represent vesicles showing invagination at the poles followed by shape instabilities. See Videos S1 and S2 in the Supplementary Material for full dynamics of the shape transformations.}
\label{fig_pd}
\end{figure}

The experimental phase diagram for vesicle shapes in a uniform AC electric field with amplitude in the range of 0-20 kV/m and a frequency at the oblate-prolate transition is shown in Figure \ref{fig_pd}. Each DOPC vesicle was observed for 60 seconds, longer than the inverse growth rate of dynamic interfacial instabilities predicted by linear stability analysis \cite{sens2002undulation,seiwert2012stability,seiwert2013instability}. Hence, it can be safely assumed that vesicles sufficiently explored the dynamics and the shapes are stationary. 

We observe that the vesicle shape is sensitive to the initial membrane tension (measured in the absence of electric field). At high membrane tension $\Sigma \sim~10^{-7}-10^{-6}$ N/m, increasing the electric field strength to 20 kV/m did not induce any morphological changes and the vesicles fluctuated about their quasi-spherical shapes at the critical frequency. Such vesicles are referred herein as stable vesicles. At significantly lower tension $\Sigma \sim~10^{-8}-10^{-9}$ N/m, the vesicles exhibited stable shapes up to a field magnitude of 10 kV/m, with recorded decrease in shape fluctuations (see Fig. \ref{figspectrume}b). However, above 10 kV/m, vesicle shapes transitioned into spindle-like configurations. Shape transitions from quasi-spherical to spindle-like shapes occurred on a time scale of 20-30 seconds; an example of spindle formation from quasi-spherical vesicles is shown in Video S1 of the Supplementary Material. We also observed that non-spherical vesicles aligned their major axis of deformation along the field direction similar to electrically-driven prolate shapes. The vesicles studied experimentally sustained their shapes during each observation window for electric field strengths up to $20$ kV/m. Turning off the field led to recovery of the initial quasi-spherical configurations and hence the shapes reported in Fig.~\ref{fig_pd} are reversible.
Tensionless GUVs with membrane tension $\Sigma <10^{-9}$ N/m dimpled at the poles upon increasing the field strength above 15 kV/m. Further field increase led to the growth of shape instabilities where the vesicles transformed into two transient and connected spindle shapes. We observed that one of the spindle-shaped compartment would grow randomly and the other would shed away reducing the excess area and increasing the membrane tension. The evolution of the shapes and instabilities described above for initially tensionless giant vesicles is shown in Video S2 of the Supplementary Material.

In general, the above observations depend on electric field strength, membrane tension, bending rigidity, and GUV size. Two dimensionless numbers describe the importance of these variable parameters: electrical capillary number, 
\begin{equation}
    \label{eq electric Ca}
    Ca_{el}=\frac{\epsilon\,E_0^2a^3}{\kappa}\,,
\end{equation}
and the reduced volume (or excess area, $\Delta$).

The electric capillary number compares shape preserving bending stresses to shape distorting electrical stresses, where $\epsilon$ is the electric permittivity and $E_0$ is the electric field magnitude. Previously, Dahl et al. \cite{dahl2016experimental} determined excess area by measuring contour fluctuations. However, as pointed out by Zhou et al.\cite{zhou2011stretching}, the fluctuations in the azimuthal direction are ignored in this method and the excess area is approximated. We circumvent this problem by directly pulling out all the fluctuations in  ellipsoidal deformation by applying uniform AC field with an increasing amplitude. The excess area is calculated from the highest aspect ratio  \cite{Hammad2022}. The relation between excess area and reduced volume \refeq{eq reduced volume} is
\begin{equation}
    \label{eq excess area}
    \nu =\left(1+\frac{\Delta}{4 \pi}\right)^{-3/2} \,.
\end{equation}

The results shown in Fig.~\ref{fig_pd} indicate three characteristic shapes in the ($\Delta$, $Ca_{el}$) phase space: i) stable vesicles, where the vesicle shape fluctuates around a mean quasi-spherical contour, for $Ca_{el}<10^4$ and excess area $0-2$; ii) spindle-like vesicles for $Ca_{el}>10^4$ and excess area between $0.2-1.2$; and iii)  unstable vesicle shapes with dimpled regions at the poles for $Ca_{el}>10^4$ and excess area greater than $0.5$.

\subsection{Electric field increases tension}
Flickering spectroscopy \cite{faizi_sm2020} of the vesicles in the absence and presence of an electric field indicates that the tension increases. In brief, the method analyses a time series of vesicle contours in the focal plane (the equator of the quasi-spherical vesicle). The quasi-circular  contour is decomposed in Fourier modes, $r(\phi)=a\left(1+\sum_q u_q(t)\exp(i q \phi)\right)$. The  fluctuating amplitudes $u_q$ are independent and have mean square amplitude  dependent only  on the membrane bending rigidity $\kappa$ and the tension $\sigma$
\begin{equation}
\langle \left|u_{q}\right |^2\rangle\sim\frac{k_BT}{\kappa\left( q^3+ \bar\Sigma q^2\right)}
\label{seq Helfrich spectrumq}
\end{equation}
where, $k_BT$ is the thermal energy, $\kappa$ is the bending rigidity, $\Sigma=\sigma R^2/\kappa$ is the dimensionless membrane tension, $q$ is the mode number, and \textbf{q}=($q_x,\,q_y$) is the wave vector conjugate to position $(x,\,y)$, where $x$ is the direction along the membrane undeformed plane, and $y$ is the direction normal it. In real space, GUV equatorial fluctuations were measured from an average value, $a_0$. Hence, we average out the theoretical
spectrum \refeq{seq Helfrich spectrumq} in the $q_y$ direction to get $\langle \left|u_{q}\right |^2\rangle\sim\frac{k_BT}{\kappa \left(q^3+\bar \Sigma q\right)}$ \cite{gracia2010effect, faizi2019bending,faizi_sm2020}. The low modes ($q<\sqrt{\bar\Sigma}$) are dominated by the tension.

\begin{figure*}
\begin{picture}(250,125)(-15,-2)
\put(-153,-1){\includegraphics[scale=0.26]{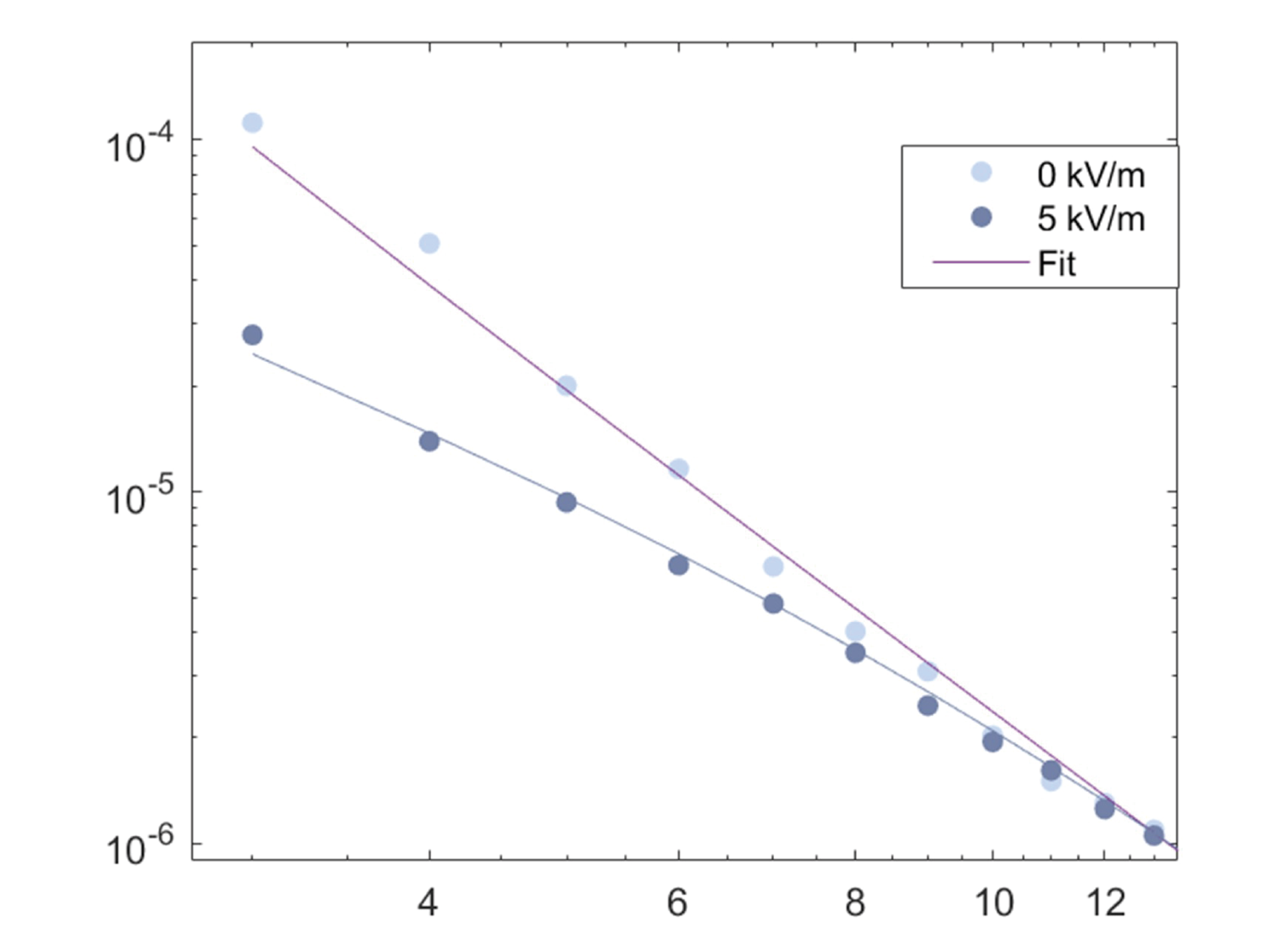} 
\put(-192,78){\large $\langle \left|u_{q}\right |^2\rangle$}
\put(-30,125){(a)}
\put(-85,-3){\large $q$}
}
\put(21,5){\includegraphics[scale=0.25]{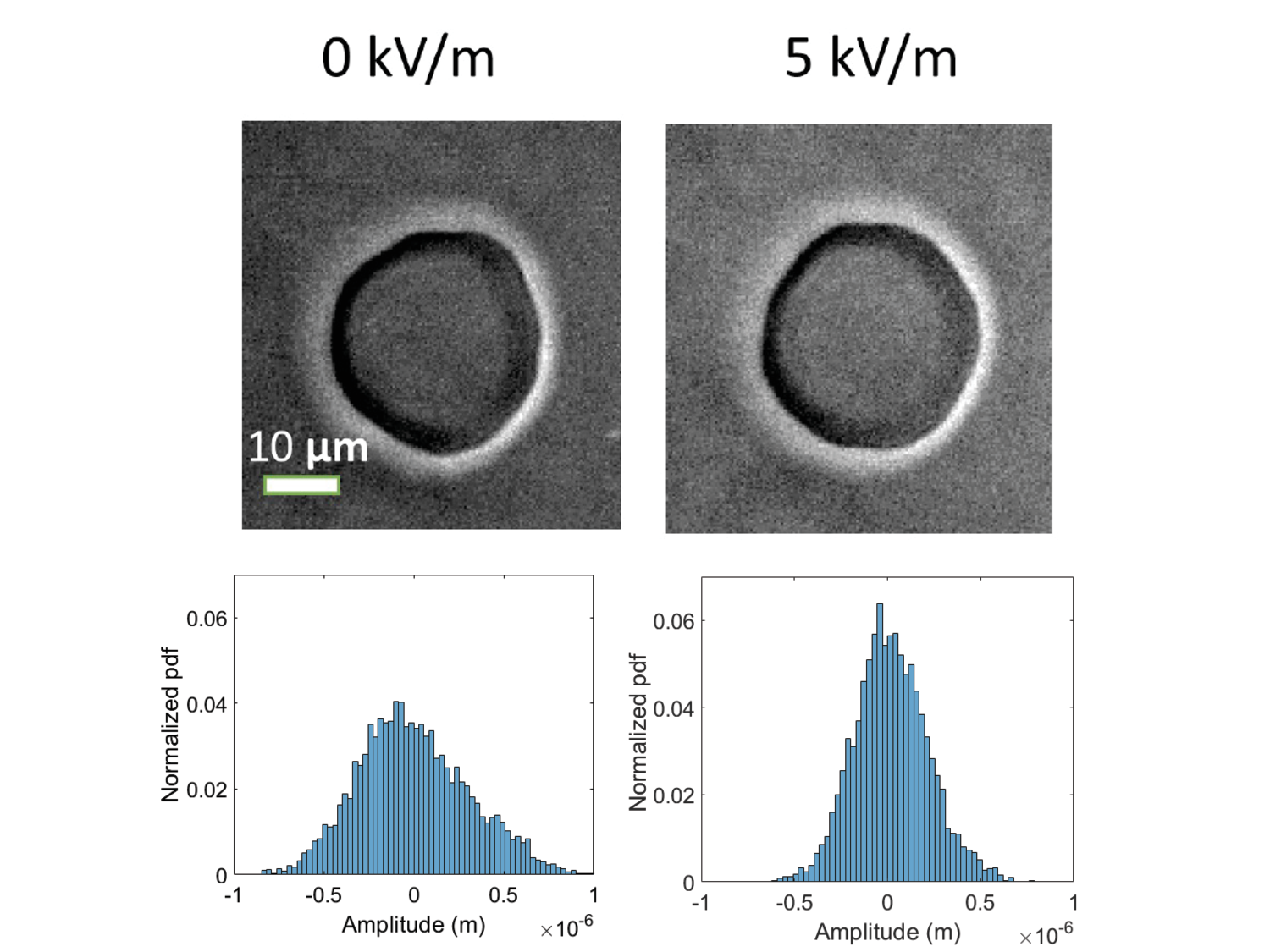} 
\put(-30,120){(b)}
}
\put(192,5){\includegraphics[scale=0.25]{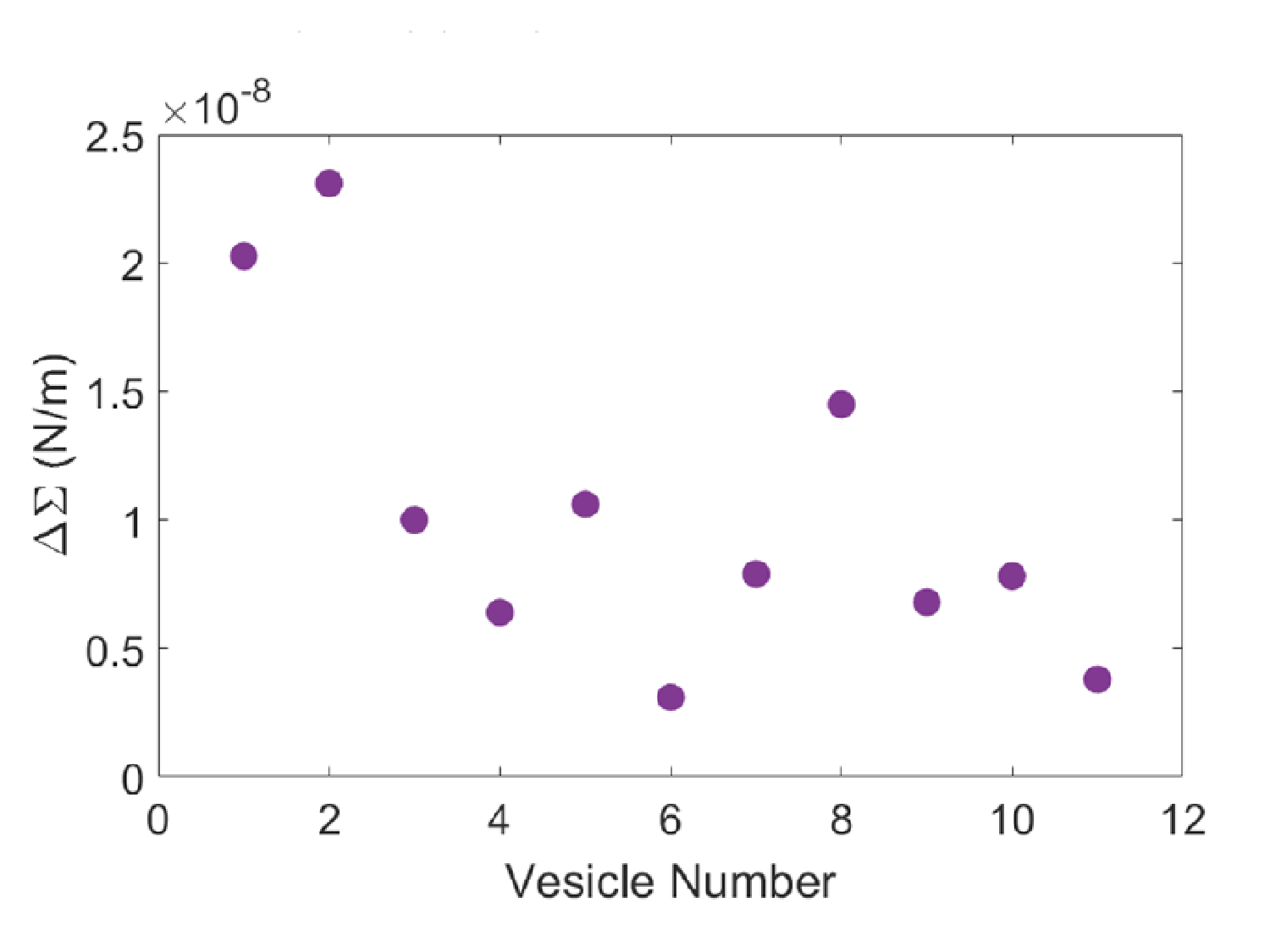}
\put(-30,105){(c)}
}
\end{picture}
\caption{a) Experimental shape fluctuation spectra as a function of mode number $q$ for GUVs as a function of the electric field strength. The solid lines are obtained from the theoretical fit of Helfrich's spectrum \refeq{seq Helfrich spectrumq} yielding membrane tension. b) Images and probability distribution of fluctuation amplitude of the vesicles at two different electric fields for part (a). c) Increase in membrane tension for an ensemble of eleven different vesicles as the electric field is switched from 0 to 5 kV/m. The inner and outer solution salt concentration are 0 mM NaCl and 1 mM NaCl respectively.}
\label{figspectrume}
\end{figure*}
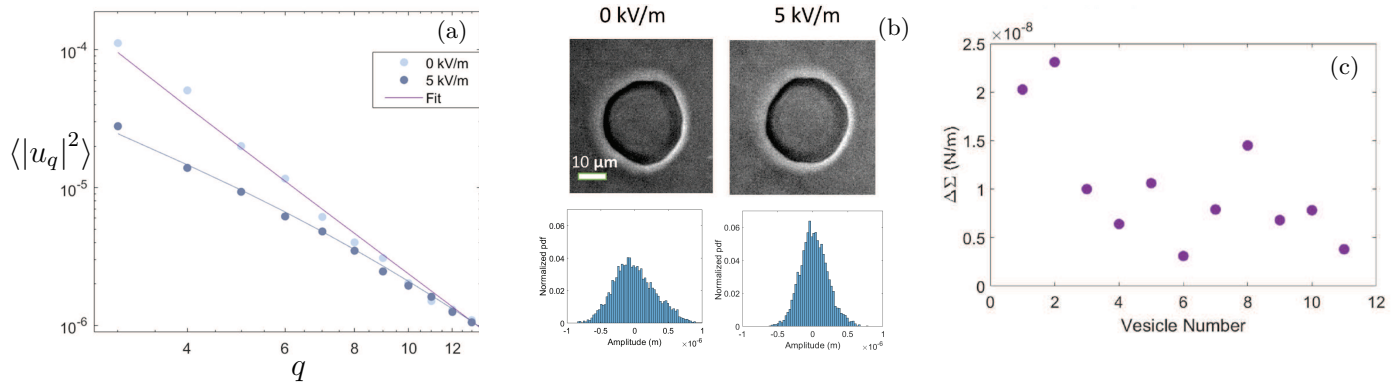

Figure \ref{figspectrume} shows the change in fluctuation spectrum \refeq{seq Helfrich spectrumq} as the applied field magnitude increases from 0-5 kV/m. The overall spectrum shift illustrated in \ref{figspectrume}(a) indicates a decrease in fluctuations following an increase in field strength. The overall decrease in amplitude of large wavelength (low wavenumber) fluctuations is shown in the probability density function (\textit{pdf}) plots of Fig.~\ref{figspectrume}(b). Quantitatively, this decrease can be characterized by the root mean square displacement (\textit{rmsd}) of edge fluctuations, $\sigma_h$, which is equivalent to the standard deviation of a Gaussian distributed histogram of the fluctuation amplitude \cite{rodriguez2015direct,betz2009atp}. The \textit{rmsd} is independent of dynamics which averages the fluctuation amplitude over time and is independent of viscosity for an equilibrated system. Without any applied electric field strength, the recorded \textit{rmsd} for membrane fluctuations were $\sigma_h=310\pm42$ nm ; on increasing the electric field strength, the fluctuations decreased to $\sigma_h=198\pm9$ nm. Using Fourier modes as a proxy for the GUVs' microscopic configurations, we find the presence of a zero probability flux between different modes using detailed balance \cite{kokot2022spontaneous}. Details are given in appendix \ref{Appendix: Detailed Balance} indicating that fluctuations are thermally driven even in the presence of electric field.  Figure \ref{figspectrume}(c) shows an increase in membrane tension with electric field strength obtained by fitting the experimental data to Eq.~\refeq{seq Helfrich spectrumq} for 11 different vesicles which is associated with an overall decrease in membrane fluctuation with electric field strength as depicted in Figs.~\ref{figspectrume}(a)-(b). 

\subsection{Comparison of experimental and \\ theoretical results}
\vspace{-.42cm}
Spindle-like configurations are observed in the numerical results shown in Fig.~\ref{fig:bending energy positive curvature} for reduced volumes in the range $0.75\lesssim \nu <1$ represented by filled circles; see, for instance, the shape in Fig.~\ref{fig:Energy Fixed A and L, variable V}(c). This range of reduced volumes encompass the experimental range of reduced volumes reported in Fig.~\ref{fig_pd} where spindle shapes are verified, i.e., $0.87 \lesssim \nu <1$ (or excess area between $0<\Delta \lesssim 1.2$). Inspection of Fig.~\ref{fig_pd} indicates an overlapping region of filled red circles and red diamonds where spindle and dimpled shapes may coexist for reduced volumes $\nu \gtrsim 0.9$; this range of reduced volumes is in qualitative agreement with the region of coexisting spindle and dimpled shapes shown in the three-dimensional phase space of Fig.~\ref{fig:3D energy curvature}. Moreover, numerical results of Fig.~\ref{fig:pressure tension positive c1} show that spindle configurations are driven by a significant increase in membrane tension for reduced volumes near one. This is in qualitative agreement with electric-field driven increase in membrane tension followed by a decrease in overall shape fluctuations and increase in field strength (cf., Figs.~\ref{figspectrume}(b) and (c)). Shape transformations from quasi-spherical vesicles at the critical frequency are reported for high values of electric capillary number $Ca_{el}>10^4$ which can be reinterpreted as a threshold of high values of critical dimensionless electric tension $\Sigma^{el} a^2/\kappa$ at which spindle-like and/or dimpled shapes may be verified. The black filled circles in Figs.~\ref{fig:bending energy positive curvature} and \ref{fig:pressure tension positive c1} indicate the range of reduced volumes contemplated by the experimental results of Fig.~\ref{fig_pd}, and the red filled circles represent the additional range of reduced volumes where stationary spindle-like shapes are numerically identified according to the assumptions presented in Section \ref{shapes with positive curvature}.

\section{Conclusions}
\label{conclusions}
In this work we developed a numerical and experimental study on stationary vesicle configurations driven by the combined effect of isotropic stresses and localized forces to interrogate membrane activity and induced mechanics of biological and synthetic cells. We show that classical results for unconstrained vesicles \cite{SeifertBerndlLipowsky1991} correspond to an envelope of lowest-energy configurations which can be driven to higher-energy stationary shapes by modulations of pressure, tension, and axial forces as analogs to externally applied fields. Numerical results reveal limiting shapes of vesicles showing spindle-like configurations and further tethering at the poles as the vesicle is pressurized or ``inflated". In the other direction, when the vesicle is ``deflated", bifurcations in the energy diagram show multi-lobed vesicle contours of increasing number of modes. Typically, spindle-like shapes are reported for vesicle reduced volumes near the spherical limit, and numerical results indicate that spindle shapes occur when the internal pressure of the vesicle exceeds the external pressure. We propose a numerical methodology that identifies a finite region in a vast parameter space of possible solutions, where stationary spindle-like shapes are identified. We further interpret our numerical results in the context of electric fields and show qualitative agreement with spindle shapes observed experimentally when giant vesicles in uniform AC fields are exposed to a broad range of electric field strengths. The results of this work elucidate some of the theoretical, numerical, and experimental challenges related to the modelling and assay of unconstrained or constrained, axisymmetric giant vesicles. Our analysis can be extended to a broader set of higher-energy states by allowing, for example, asymmetries in the packing of the lipid molecules (i.e., $C_0\neq 0$), and spatial variations of membrane material properties (e.g., bending rigidity) leading to more detailed models for the bending energy density \refeq{eq bending energy density}. 

\section*{Conflicts of interest}
There are no conflicts to declare.

\section*{Acknowledgements}
This research was supported by NIGMS award \\ 1R01GM140461.

\appendix

\section{Derivation of the stationary shape equations}\label{Appendix shape equations}
For completeness, we present in this Appendix a derivation of the general shape equation \refeq{Shape Eq normal} extending the derivation for two-dimensional vesicles \cite{veerapaneni2009boundary} to three-dimensional, axisymmetric geometries. Stationary shapes are determined when the first variation of the total elastic energy \refeq{eq Bending Energy point force} is zero under small perturbations in the shape. Here, we follow the standard differential geometry notation summarized in the Supplementary Material, where the mean curvature is given by, 
$\mathrm{H}=(c_1+c_2)/2$, which differs from the definition of mean curvature appearing in Eq.~\refeq{Shape Eq normal} by a minus sign; the definition of the Gaussian curvature remains the same as in Eq.~\refeq{eq Gaussian curvature}. Assuming arclength parametrization, the perturbed shape of a vesicle can be written as
\begin{equation}
    \label{eq: perturbed shape x}
    \bar{\mathbf{x}}(s,\phi,\epsilon)=\mathbf{x}(s,\phi)+\delta \mathbf{x}(s,\phi,\epsilon)\,,
\end{equation}
where the perturbation in shape is 
\begin{equation}
    \label{eq: perturbation}
    \delta \mathbf{x}(s,\phi,\epsilon)=\epsilon\, \mathbf{y}(s,\phi)\,,
\end{equation}
and 
\begin{equation}
    \label{eq: perturbed shape y}
    \mathbf{y}(s,\phi)=u(s)\mathbf{x}_{s}(s,\phi)+v(s)\mathbf{n}(s,\phi)\,,
\end{equation}
where $\mathbf{x}_{s}$ is the tangent vector along the arclength direction $s$, $\mathbf{n}$ is the outward-pointing normal vector following the geometric convention shown in Fig.~\ref{fig:coordinate system}(b), and  $u$ and $v$ are magnitudes of perturbations in shape in the tangential and normal directions, respectively. For axisymmetric geometries, perturbations in the azimuthal direction can be neglected without loss of generality. Hence, the coefficients of the first fundamental form of the slightly deformed regular surface embedded in $\mathbb{R}^3$ are 
\begin{equation}
    \label{eq: Coeff E Ip}
    \bar E(s,\phi,\epsilon)=\bar{\mathbf{x}}_{s}\cdot \bar{\mathbf{x}}_{s}\,,
\end{equation}
\begin{equation}
    \label{eq: Coeff F Ip}
    \bar F(s,\phi,\epsilon)=\bar{\mathbf{x}}_{s}\cdot \bar{\mathbf{x}}_{\phi}\,,
\end{equation}
\begin{equation}
    \label{eq: Coeff G Ip}
    \bar G(s,\phi,\epsilon)=\bar{\mathbf{x}}_{\phi}\cdot \bar{\mathbf{x}}_{\phi}\,,
\end{equation}
where the vectors $(\bar{\mathbf{x}}_s,\bar{\mathbf{x}}_\phi)$ locally span the tangent plane at a point P on the perturbed surface where a normal vector is given by
\begin{equation}
    \label{eq perturbed normal}
    \bar{\mathbf{n}}(s,\phi,\epsilon)=\frac{\bar{\mathbf{x}}_{s}
    \times \bar{\mathbf{x}}_{\phi}}{\bar W}\,,
\end{equation}
where
\begin{equation}
    \label{eq perturbed metric}
    \bar W(s,\phi,\epsilon)=\sqrt{\bar E \bar G-\bar{F}^2} \,,
\end{equation}
defines the metric of the deformed surface \cite{struik1961lectures,doCarmo2016differential} following the derivation in the Supplementary Material. Accordingly, the coefficients of the second fundamental form reduce to
\begin{equation}
    \label{eq: Coeff L IIp}
    \bar L(s,\phi,\epsilon)=\bar{\mathbf{x}}_{ss}\cdot \bar{\mathbf{n}}\,,
\end{equation}
\begin{equation}
    \label{eq: Coeff M IIp}
    \bar M(s,\phi,\epsilon)=\bar{\mathbf{x}}_{s\,\phi}\cdot \bar{\mathbf{n}}\,,
\end{equation}
\begin{equation}
    \label{eq: Coeff N IIp}
    \bar N(s,\phi,\epsilon)=\bar{\mathbf{x}}_{\phi \phi}\cdot \bar{\mathbf{n}}\,.
\end{equation}

For surfaces of revolution where $z$ is the axis of symmetry in the cylindrical coordinate system ($r,\phi,z$), the vector components of the unperturbed shape are
\begin{equation}
    \label{eq cylindrical surface}
    \mathbf{x}(s,\phi)=\{r(s)\cos \phi,r(s)\sin \phi,z(s)\}\,,
\end{equation}
where $0 \le s \le L$ and $0\le \phi \le 2\,\pi$, and the normal vector reduces to
\begin{equation}
    \label{eq normal cylindrical surface}
    \mathbf{n}(s,\phi)=\{-z_s \cos \phi,-z_s \sin \phi,r_s\}\,.
\end{equation}
In this case, the lines of curvature are equal to the parametric lines reducing the expressions for the principal curvatures to 
\begin{equation}
    \label{eq principal curvature c1 Appendix}
    \bar{c}_1=\frac{\bar L}{\bar E}\,,
\end{equation}
and
\begin{equation}
    \label{eq principal curvature c2 Appendix}
    \bar{c}_2=\frac{\bar N}{\bar G}\,,
\end{equation}
as shown in the Supplementary Material.

Substituting Eqs.~\refeq{eq cylindrical surface}-\refeq{eq normal cylindrical surface} into Eqs.\refeq{eq: perturbed shape x}-\refeq{eq: Coeff N IIp} and using the arclength relation \refeq{eq arclength}, the perturbed principal curvatures \refeq{eq principal curvature c1 Appendix}-\refeq{eq principal curvature c2 Appendix} can be written as

\begin{equation}
    \label{eq principal curvature c1 Appendix 2}
    \bar{c}_1(s,\epsilon)=c_1 +\epsilon \left( v_{ss} +u\,c_{{1}_s}+v c^2_1\right)+O(\epsilon^2) \,,
\end{equation}

\begin{equation}
    \label{eq principal curvature c2 Appendix 2}
    \bar{c}_2(s,\epsilon)=c_2 +\epsilon \left(\frac{r_s}{r}v_s+v c^2_2 +u\,c_{{2}_s}\right)+O(\epsilon^2)\,.
\end{equation}
Thus,
\begin{equation}
    \label{eq mean curvature appendix}
    \bar{\mathrm{H}}(s,\epsilon)=\mathrm{H}+\epsilon \left[\frac{1}{2}\Delta_s v +v(2\mathrm{H}^2-K)+u \mathrm{H}_s\right]+O(\epsilon^2)\,,
\end{equation}

\begin{equation}
    \label{eq: W simplified}
    \bar W(s,\epsilon)=r +\epsilon \,r \left[\frac{(u r)_s}{r}-2 v \mathrm{H}\right] +O(\epsilon^2)\,,
\end{equation}
and 
\begin{equation}
    \label{eq normal bar Appendix}
    \bar{\mathbf{n}}(s,\epsilon)=\mathbf{n}(s)-\epsilon \,r \begin{bmatrix} \displaystyle \frac{r_s}{r} (c_1 \, u+v_s) \\ 0 \\ c_1 c_2 \,u +c_2 v_s \end{bmatrix} \,,
\end{equation}
where $\mathrm{H}$ and $K$ are the mean curvature and Gaussian curvature, respectively, the subscript $s$ denotes arclength derivatives, and $\bar{\mathbf{n}}$ is the perturbed normal vector defined in Eq.~\refeq{eq perturbed normal}. For presentation purposes, Eqs.~\refeq{eq principal curvature c1 Appendix 2}-\refeq{eq normal bar Appendix} are carried out using an arbitrary, fixed value for the azimuthal angle, $\phi=0$.

The first variation of the bending energy \refeq{eq Bending Energy} with $C_0=0$ is 
\begin{equation}
    \label{eq: variation bending}
    \frac{\delta^{(1)} E_b}{4\pi\kappa} =\int_{\Gamma} \left(\bar{\mathrm{H}}^2 \bar W-\mathrm{H}^2 W \right)\,ds \,,
\end{equation}
where $\Gamma$ indicates the vesicle contour where $s_1 \le s \le s_2$. Inserting Eqs.~\refeq{eq mean curvature appendix}-\refeq{eq: W simplified} into \refeq{eq: variation bending} yields
\begin{equation}
\begin{split}
    \label{eq: variation bending eps}
    \frac{\delta^{(1)} E_b}{2\pi} &=2 \kappa\epsilon \int_{\Gamma} \left[2 \mathrm{H}(\mathrm{H}^2-K)v \right. \\ 
    & \left. + u\frac{(\mathrm{H}^2 r)_s}{r} +\mathrm{H}^2 u_s + \mathrm{H}\frac{(r v_s)_s}{r}\right] r \,ds \,,
\end{split}
\end{equation}
and integrating by parts, 

\begin{equation}
\label{eq: variation bending eps B}
\begin{split}    
    \frac{\delta^{(1)} E_b}{2\pi}& = 2 \kappa \epsilon \left \{ \int^{s_2}_{s_1} \left[2 \mathrm{H}(\mathrm{H}^2-K) + \frac{1}{r}\left(\mathrm{H}_s r\right)_s\right] v\,r ds \right. \\
    & \left. \left.+ (\mathrm{H} r) v_s \right \vert^{s2}_{s_1}-\left.(\mathrm{H}_s r) v \right \vert^{s2}_{s_1} \right \}\,,
\end{split}
\end{equation}
where we assume that $\mathrm{H} r \to 0$ as $s \to 0$ and that $\mathrm{H}_s r$ is finite at the poles. This is confirmed by the local analysis presented in Appendix \ref{sec local analysis}. Hence, the first boundary term in Eq.~\refeq{eq: variation bending eps B} vanishes yielding  
\begin{equation}
    \label{eq: variation bending eps C}
\begin{split}
    \frac{\delta^{(1)} E_b}{2\pi} & = \int^{s_1}_{s_2} \left[4 \kappa \mathrm{H}(\mathrm{H}^2-K) + 2 \kappa \frac{1}{r}\left(\mathrm{H}_s r\right)_s\right] \delta \mathbf{x}\cdot \mathbf{n}\,r \,ds \\
    & -2 \kappa \left. (\mathrm{H}_s r) \delta z \right \vert^{s2}_{s_1}
\end{split}
\end{equation}
where $\delta \mathbf{x} \cdot \mathbf{n}=\epsilon v$ and $\delta \mathbf{x} \cdot \mathbf{n}\vert_{poles}=\delta z$ according to Eqs.~\refeq{eq: perturbation} and \refeq{eq: perturbed shape y}. The last term in the integrand of Eq.~\refeq{eq: variation bending eps C} is the Laplace-Beltrami operator acting on $\mathrm{H}$ for axisymmetric geometries. 

The variation of the point-force term in Eq.~\refeq{eq Bending Energy point force} depends on the sign of the mean curvature. Assuming the force is an odd function of the curvature at the poles (i.e., $F(\mathrm{H})=-F(H)$), the total elastic energy \refeq{eq Bending Energy point force} can be written as
\begin{equation}
    \label{eq: Bending Energy point force Hrm}
    \tilde{G}=\tilde{E}'(\mathrm{H},\Sigma,P)+F(\mathrm{H})\Delta z\vert^{s_2}_{s1}
\end{equation}
assuming symmetry of the forces acting at the poles, where $\tilde{E'}=E'$ for $C_0=0$.

Hence, the variation of axial force in terms of $\mathrm{H}$ is
\begin{equation}
    \label{eq: variation point force}
    \delta^{(1)} \left.( F(\mathrm{H})(z-z_0))\right\vert^{s_2}_{s_1}=F(\mathrm{H}) \delta z\vert^{s_2}_{s_1}\,.
\end{equation}

Similarly, the first variation of the tension term in \refeq{eq Bending Energy no force} can be written as 
\begin{equation}
    \label{eq: variation tension eps}
    \begin{split}
    \frac{\delta^{(1)} E_{\Sigma}}{2\pi} & = \int_{\Gamma} \Sigma \left(\bar W - W\right) \,ds \\ &=\epsilon \int_{\Gamma} \Sigma  \left[\frac{(u r)_s}{r}-2 v \mathrm{H}\right]\,r\,ds\,,
    \end{split}
\end{equation}
and integration by parts yields
\begin{equation}
    \label{eq: variation tension eps B}
    \frac{\delta^{(1)} E_{\Sigma}}{2\pi} = -\int_{\Gamma} \nabla_s \Sigma\, \delta \mathbf{x}\cdot \mathbf{x}_s \,r \,ds  - \int_{\Gamma} \left(2 \mathrm{H} \Sigma \right) \delta \mathbf{x}\cdot \mathbf{n}\, r \,ds\,,
\end{equation}
where the boundary term $(\Sigma r) u \vert^{s_2}_{s_1}$ vanishes for a finite tension. The $O(\epsilon)$ first variation of the volume is
\begin{equation}
    \label{eq: variation pressure eps A}
    \frac{\delta^{(1)} E_{P}}{2\pi} =  P  \int_{\Gamma}  \frac{1}{3} \left[\bar{W}(\bar{\mathbf{n}}\cdot \bar{\mathbf{x}})- W (\mathbf{n}\cdot \mathbf{x})\right] \,ds,
\end{equation}
where the differential volume is defined as  
\begin{equation}
    \label{eq dV diff geometry}
    \int_{V} dV = \frac{1}{3}\int_{\partial V} dA \,\mathbf{n}\cdot \mathbf{x} \,,
\end{equation}
and the differential area is given by $dA=\sqrt{EG-F^2} du\,dv$ (see Supplementary Material). Using relations \refeq{eq: W simplified}-\refeq{eq normal bar Appendix} and definitions \refeq{eq: perturbed shape x}-\refeq{eq: perturbed shape y} and integrating by parts yields
\begin{equation}
    \label{eq: variation pressure eps B}
    \frac{\delta^{(1)} E_{P}}{2\pi} =  P \int_{\Gamma}  \delta \mathbf{x}\cdot \mathbf{n} \,r \,ds,
\end{equation}
where $P$ is the difference between exterior and interior pressures across the membrane. 

Setting to zero the total variation of the total elastic energy \refeq{eq Bending Energy point force} with respect to arbitrary perturbations in shape yields 
\begin{equation}
    \label{Shape Eq normal tilde}
    2 \kappa \Delta_{b} \mathrm{H}+4 \kappa \mathrm{H}(\mathrm{H}^2-K)-2 \mathrm{H} \Sigma +P = 0 \,,
\end{equation}
and the relation 
\begin{equation}
    \label{eq: force relation}
    F=4\pi \kappa (\mathrm{H}_s r)\,,
\end{equation}
for the point force at the poles. 

Letting $\mathrm{H}=-H$ following the notation in Refs. \cite{SeifertBerndlLipowsky1991,Seifert1997} recovers the shape equation \refeq{Shape Eq normal} and the axial force balance \refeq{eq axial force H}. Note that the minus sign applies to the definition of the mean curvature, and the definition of the principal curvatures remains the same as in Eqs.~\refeq{eq meridional curvature} and \refeq{eq azimuthal curvature}.

\section{Pseudo-spectral numerical solution}\label{Appendix pseudospectral}
In this Appendix, we present details of the pseudo-spectral numerical solution of Eqs.~\refeq{eq Shape dimensionless normal}-\refeq{eq arclength dimensionless} and boundary conditions \refeq{boundary condition r}-\refeq{boundary condition z}. Here, the characteristic length scale is $l_c=L$, and the solution domain is mapped onto the interval $-1\le \hat x\le 1$. Using a change of variable,
\begin{equation}
    \label{eq collocation change of variables}
     \hat x =1-2 \hat s \,,
\end{equation}
the governing equations and boundary conditions reduce to
\begin{equation}
\begin{split}
    \label{eq Shape dimensionless normal hat}
    &-\left[\hat r (4 \hat c_1+\hat c_2)_{x}\right]_{x}-\frac{\hat r}{2}\left(4 \hat c_1+\hat c_2\right)\left(4 \hat c_1-\hat c_2\right)^{2} \\ 
    & +\frac{\hat \Sigma}{\hat \kappa}\frac{\hat r}{4}\left(4 \hat c_1+\hat c_2\right)-\frac{\hat r}{8}\frac{\hat P}{\hat \kappa}= 0\,,
\end{split}
\end{equation}

\begin{equation}
    \label{eq arclength dimensionless hat}
   (\hat r_x)^2 +(\hat z_x)^2 = \frac{1}{4}  \,,
\end{equation}

\begin{equation}
    \label{boundary condition r hat}
    \hat r(0)=0\,, \qquad \hat r(1)=0\,,
\end{equation}

\begin{equation}
\label{boundary condition z hat}
    \hat z_s(0)=0\,, \qquad \hat z_s(1)=0\,,
\end{equation}
where the over-hat superscript denotes variables normalized by $L$. The computational domain is discretized in $N+1$ Chebyshev collocation points,
\begin{equation}
    \label{eq Chebyshev points}
    \hat{x}_i =\cos \left(\frac{(i-1) \pi}{N}\right)\,, \qquad i=1,\dots,N+1\,, 
\end{equation}
and the resulting system of non-linear algebraic equations is solved using Newton's method. The results shown in section \ref{sec:results} were computed with $N=75$ where spectral accuracy is verified. 

\subsection{Numerical convergence}
\begin{figure*}
\begin{picture}(250,120)(-15,0)
\put(-135,5){\includegraphics[scale=0.5]{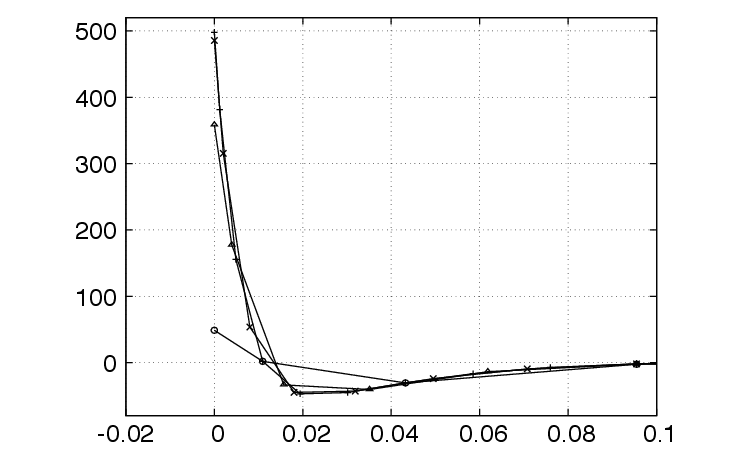} 
\put(-121,78){\large $\delta/L=0.025$}
\put(-45,90){(a)}
\put(-185
 ,60){\large $\hat r_{sss}$}
\put(-90,-12){\large $\hat s$}
}
\put(20,5){\includegraphics[scale=0.5]{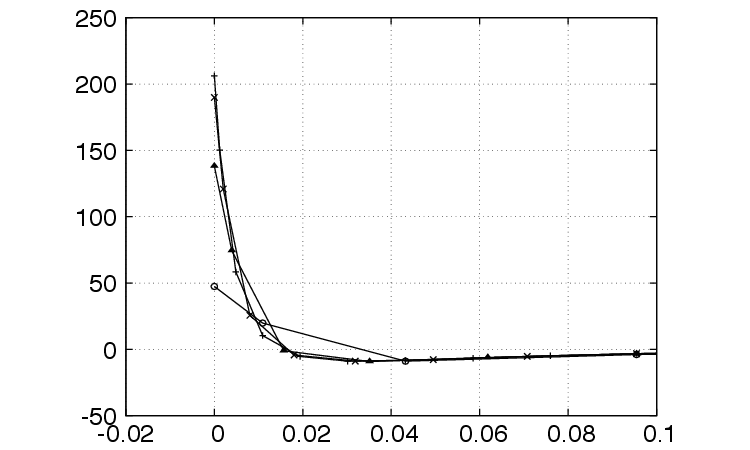} 
\put(-45,90){(b)}
\put(-121,78){\large $\delta/L=0.015$}
\put(-90,-12){\large $\hat s$}
}
\put(175,5){\includegraphics[scale=0.5]{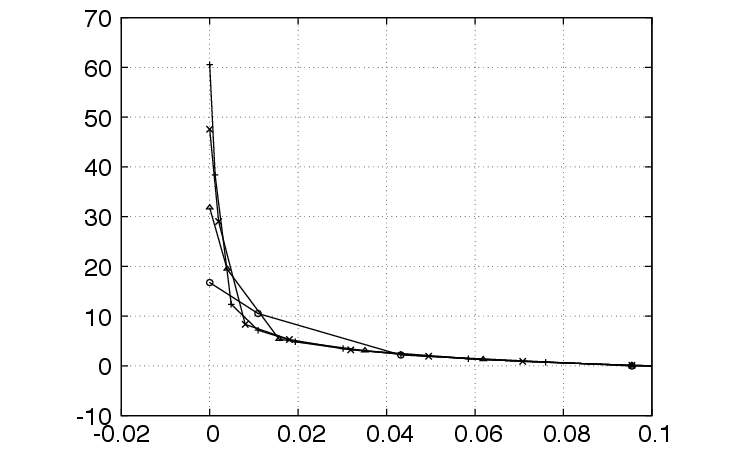}
\put(-121,78){\large $\delta/L=0.005$}
\put(-45,90){(c)}
\put(-90,-12){\large $\hat s$}
}
\end{picture}
\caption{Numerical results for $\hat r_{sss}$ versus arclength using $l_c=L$ for three difference circular gaps at the poles indicated by $\delta/L$.}
\label{fig: Numerical convergence}
\end{figure*}
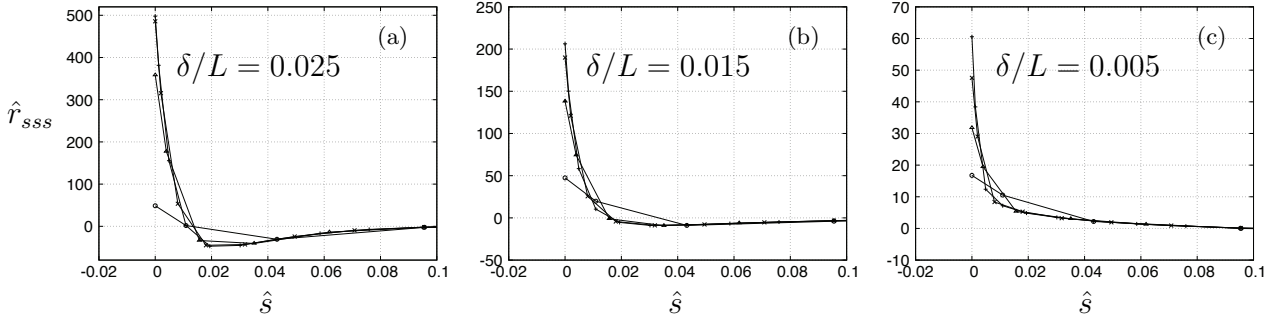

We present numerical results for the third-derivative of the spatial variable $r$ for a hypothetical, opened vesicle shape at the poles as an example of numerical convergence of our results away from the singular limit of closed, constrained vesicles at $s=0$. Equations.~\refeq{eq Shape dimensionless normal}-\refeq{eq arclength dimensionless} are solved with modified boundary conditions $\hat r(0)=\hat r(1)=\delta/L$ and $\hat z_s(0)=\hat z_s(1)=0$ for $N=45,35,25,15$ (top to bottom curves) shown in Fig.~\ref{fig: Numerical convergence}(a)-(c); values for the circular spacing at both ends of the vesicle are indicated in the figure. Each plot (a)-(c) is obtained by fixing the area and length of the vesicle, and changing the volume. This is equivalent to moving up the solid-green curve in Fig.~\ref{fig:Energy Fixed A and L, variable V} starting from a reduced volume of 0.90 by volume increments  $\nu=\nu \,d\nu$ where $d\nu=(1.005^2)$ is the incremental change. Exponential convergence is verified as the number of points increases, as indicated by the relative distance between pairs of points for a fixed arclength.

\section{Tilt-angle formulation for axisymmetric vesicles}\label{sec: Euler-Lagrange discussion}
The tilt-angle formulation has been extensively used in numerical analyses of stationary shapes of axisymmetric vesicles \cite{Seifert1997}, where the tilt angle $\psi$ is subtended between the tangent vector to the surface and the horizontal direction, as illustrated in Fig.~\ref{fig:coordinate system}(b). In this Appendix, we revisit this formulation following the derivation presented in Refs.~\cite{SeifertBerndlLipowsky1991,JulicherSeifert1994} for completeness of presentation and further comparison with the numerical results shown in section \ref{sec:results}. First, a derivation of the shape equations assuming constant-force ensemble is presented and comments are made about the dynamically equivalent shape equations assuming constant-height.

The general shape equation \refeq{Shape Eq normal} can be recast as a system of non-linear ordinary differential equations for axisymmetric vesicles by minimizing the total energy functional \refeq{eq Bending Energy point force}. In the tilt-angle approach, the variables $(r,z,\psi;r_s,z_s,\psi_s)$ are taken as independent ``coordinates" and ``velocites", where arclength plays the role of time in classical mechanics; thus, the geometric relations between the spatial coordinates $(r,z)$ and the tilt angle 
\begin{equation}
    \label{eq drds dzds tilt}
    r_s=\cos \psi\,, \quad z_s=-\sin \psi \,,
\end{equation}
are enforced via Lagrange multipliers $(\gamma,\eta)$, respectively, where $\psi \in [0,\pi]$ for $0 \leq s \leq L$. The total elastic energy \refeq{eq Bending Energy point force} can be written in the terms of a ``Lagrangian" function, $\mathcal{L}$, as follows
\begin{equation}
    \label{eq energy functional A}
    G'_{\mathcal{L}}=2 \pi \kappa \int_{s_1}^{s_2} \mathcal{L}(r,r_s,z_s,\psi,\psi_s) \, ds - F \Delta z \vert_{s=s_1}\,,
\end{equation}
where $s_1$ and $s_2$ are the arclength measures at the north and south poles, respectively, and
\begin{equation}
    \label{eq Lagrangian}
\begin{split}
     \mathcal{L}&=\frac{r}{2}\left(\psi_s+\frac{\sin \psi}{r}\right)^2+\frac{\Sigma}{\kappa}r+\frac{1}{2}\frac{P}{\kappa} r^2 \sin \psi \\ 
     &+\gamma(r_s-\cos \psi) + \eta (z_s +\sin \psi) \,.
\end{split}
\end{equation}
In Eq.~\refeq{eq Lagrangian}, the membrane is assumed symmetric and the principal curvatures are given by 
\begin{equation}
    \label{eq principal curvatures tilt angle}
    c_1=-\psi_s \,,   \quad c_2=-\frac{\sin \psi}{r}\,,
\end{equation}
as shown in Appendix \ref{Appendix shape equations} and given by definitions \refeq{eq mean curvature}, \refeq{eq meridional curvature}, \refeq{eq azimuthal curvature}, and the geometric relations \refeq{eq drds dzds tilt}.

Following Halminton's principle of stationary action derived in the Supplementary Material for completeness, where the action functional is given by Eq.~\refeq{eq energy functional A} and the arclength $s$ is treated as time, extrema conditions on the membrane elastic energy are obtained by path variations of the energy functional in the configurational space spanned by the coordinates $(r,z,\psi)$. Combining Eqs.~(2.3) and (2.6) in the Supplementary Material, the variation of the elastic energy reduces to
\begin{equation}    
     \label{eq variation of F'}
     \begin{split}
    \frac{\delta G'_{\mathcal{L}}}{2\pi \kappa} &=\int_{s_1}^{s_2}  \left\{ \left[\frac{\partial \mathcal{L}}{\partial \psi} -\frac{d}{ds}\frac{\partial \mathcal{L}}{\partial \psi_s}\right]\delta \psi  + \left[\frac{\partial \mathcal{L}}{\partial r} -\frac{d}{ds}\frac{\partial \mathcal{L}}{\partial r_s}\right]\delta r \right.\\
     & \left. + \left[\frac{\partial \mathcal{L}}{\partial z} -\frac{d}{ds}\frac{\partial \mathcal{L}}{\partial z_s}\right]\delta z \right\} ds-\mathcal{H} \Delta s\lvert_{s_1}^{s_2} +\left.\frac{\partial \mathcal{L}}{\partial \psi_s }\Delta \psi \right|_{s_1}^{s_2} \\
     &+\left.\frac{\partial \mathcal{L}}{\partial r_s }\Delta r\right|_{s1}^{s2}+\left.\frac{\partial \mathcal{L}}{\partial z_s }\Delta z\right|_{s_1}^{s_2} -\frac{F}{2 \pi \kappa} \Delta z|_{s_1} \,,
     \end{split}
\end{equation}
where $\delta E'_{\mathcal{L}}=0$ gives a stationary shape, and variations of each coordinate at the poles are given by 
\begin{equation}
    \Delta r=\delta r+r_s \Delta s\,,\quad  \Delta z=\delta z+z_s \Delta s\,,\quad \Delta \psi=\delta \psi+\psi_s \Delta s \,,
\end{equation}
according to (2.5) in the Supplementary Material. In Eq.~\refeq{eq variation of F'}, $\mathcal{H}\equiv \mathcal{L}-\psi_s \partial \mathcal{L}/\partial \psi_s-r_s \partial \mathcal{L}/\partial r_s -z_s \partial \mathcal{L}/\partial z_s $ plays the role of the Hamiltonian of the system
\begin{equation}
    \label{eq Hamiltonian}
    \begin{split}
     \mathcal{H}&=\frac{r}{2}\left[\psi^2_s-\left(\frac{\sin \psi}{r}\right)^2\right]-\frac{\Sigma}{\kappa}r-\frac{1}{2}\frac{P}{\kappa} r^2 \sin \psi \\
     &+\gamma \cos \psi - \eta \sin \psi \,,
     \end{split}
\end{equation}
as defined in Eq.~(2.7) in Supplementary Materials. Given that the Langrangian function \refeq{eq Lagrangian} is not an explicit function of arclength, i.e. $\frac{\partial \mathcal{L}}{\partial s}=0$, then
\begin{equation}
    \label{eq condition H const}
   \frac{d\mathcal{H}}{d s}=0 
\end{equation}
and $\mathcal{H}$ is constant.    

When $\delta G'_{\mathcal{L}}=0$, the terms in the integrand of Eq.~\refeq{eq variation of F'} yield a system of Euler-Lagrange shape equations for arbitrary variations of $(r,z,\psi)$ as follows, 
\begin{equation}
    \label{eq JulicherSeifert 1}
    \begin{split}
     \psi_{ss}=&\frac{\cos \psi \sin \psi}{r^2}-\frac{\psi_s}{r}\cos \psi +\frac{1}{2}\frac{P}{\kappa} r \cos \psi \\ & +\frac{\gamma}{r}\sin \psi +\frac{\eta}{r} \cos \psi \,,
     \end{split}
\end{equation}

\begin{equation}
    \label{eq JulicherSeifert 2}
     \gamma_{s}=\frac{1}{2}\psi^2_s -\frac{\sin^2 \psi}{2\,r^2}+\frac{\Sigma}{\kappa}+\frac{P}{\kappa} r \sin \psi \,,
\end{equation}

\begin{equation}
    \label{eq JulicherSeifert 3}
     \eta_{s}=0 \,,
\end{equation}
where the Lagrange multiplier functions $\gamma$ and $\eta$ enforce the geometric arclength relation \refeq{eq arclength} locally. 

Boundary conditions \refeq{boundary condition r}-\refeq{boundary condition z} still apply for the system of equations \refeq{eq JulicherSeifert 1}-\refeq{eq JulicherSeifert 3} and the geometric relations \refeq{eq drds dzds tilt}. Note that Eq.~\refeq{boundary condition z} combined with the arclength relation \refeq{eq arclength} and the radial geometric constraint (i.e., $r_s=\cos \psi$) yield equivalent boundary conditions for the tilt angle,
\begin{equation}
\label{boundary condition psi }
    \psi(0)=0\,, \qquad \psi(L)=\pi\,,
\end{equation}
which enforce that the first three terms and the last term in Eq.~\refeq{eq Hamiltonian} vanish, leading to 
\begin{equation}
\label{boundary condition gamma }
\gamma(0)=\gamma(L)=\mathcal{H}\,,
\end{equation}
where,
\begin{equation}
    \label{eq Hamiltonian eq 0}
    \mathcal{H}\equiv 0\,,
\end{equation}
for arbitrary variations in arclength at the poles.

Constraints of constant area and constant volume can be imposed globally using
\begin{equation}
    \label{eq area}
     A_T - \int^{L}_{0} 2\pi r \,ds=0\,,
\end{equation}
and
\begin{equation}
    \label{eq volume}
     V_T - \int^{L}_{0} \pi r^2 \sin \psi ds=0\,.
\end{equation}

Equation \refeq{eq Hamiltonian eq 0} implies that the length $L$ of the vesicle is determined self-consistently (i.e., for $\Delta s\vert_{poles} \neq 0$) to satisfy  the extremum condition on the elastic energy, $\delta E'_{\mathcal{L}}=0$. Moreover, boundary conditions of fixed angles at the poles and closed vesicle shapes yield $\Delta \psi\vert_{poles}=\Delta r \vert_{poles}=0$, respectively. For non-zero changes in height of the vesicle, $\Delta z\vert_{s_1}\neq 0$, a point force 
\begin{equation}
    \label{eq remaining term dE=0}
    F=2 \pi \kappa \eta \,,
\end{equation}
is needed to enforce $\delta E'_{\mathcal{L}}=0$, where we assumed, by symmetry, that the forces acting on both poles are equal and point in opposite directions. The same relation for the force \refeq{eq remaining term dE=0} is recovered in Appendix \ref{sec local analysis} using the local behavior of the tilt angle \refeq{eq tilt sol local s=0}, the extremum of the energy \refeq{eq Bending Energy point force}, and the definition of the axial force \refeq{eq axial force H}.

The system of Euler-Lagrange equations~\refeq{eq JulicherSeifert 1}-\refeq{eq JulicherSeifert 3} and boundary conditions \refeq{boundary condition r}, \refeq{boundary condition psi }, \refeq{boundary condition gamma }, and \ref{eq Hamiltonian eq 0} along with the geometric relations \refeq{eq drds dzds tilt}, \refeq{eq area} and \refeq{eq volume} can be solved numerically for axisymmetric vesicle shapes. A possible numerical approach is to use an implicitly, two-point boundary value problem in a truncated domain with modified boundary conditions to avoid coordinate singularities at the poles \cite{powers2007vesicle}. This analysis can be conducted for an ensemble of axisymmetric membranes with edges (or holes at both poles) hold at a fixed separation by an axial force, where the same form of Eq.~\refeq{eq remaining term dE=0} has been derived in Ref. \cite{jia2021axisymmetric}.

In the constant-height scenario, the potential \refeq{eq energy functional A} is modified using relation \refeq{eq remaining term dE=0} directly,\cite{bozic1997theoretical,derenyi2002formation}
\begin{equation}
    \label{eq energy functional h0}
    \tilde{G}'=2 \pi \kappa \int_{s_1}^{s_2} \mathcal{\tilde{L}}(r,r_s,\psi,\psi_s) \, ds \,,
\end{equation}
where the axial force appears in the modified Lagrangian and enforces the geometrical constraint of constant height as follows
\begin{equation}
    \label{eq Lagrangian const h0}
    \begin{split}
     \mathcal{\tilde{L}}&=\frac{r}{2}\left(\psi_s+\frac{\sin \psi}{r}\right)^2+\frac{\Sigma}{\kappa}r+\frac{1}{2}\frac{P}{\kappa} r^2 \sin \psi \\
     &+\gamma(r_s-\cos \psi) + \frac{F}{2 \pi \kappa} \sin \psi \,,
     \end{split}
\end{equation}
and
\begin{equation}
    \label{eq height constraint h0}
    h_0+\int^{L}_{0} \sin \psi ds =0 \,,
\end{equation}
using $z_s=-\sin \psi$. Taking the first variation of \refeq{eq energy functional h0} following the steps used in the energy minimization of Eq.~\refeq{eq variation of F'}, yields a dynamically equiavlent system of Euler-Lagrange equations \refeq{eq JulicherSeifert 1}-\refeq{eq JulicherSeifert 2} where one uses Eq.~\refeq{eq remaining term dE=0} to eliminate $\eta$. In this case, the boundary conditions are $r(0)=r(L)=0$, $z(0)=0$ and $z(L)=h_0$ with the geometric constraints on area, volume, and height given by \refeq{eq area}, \refeq{eq volume}, and \refeq{eq height constraint h0}, respectively. The boundary terms in the energy minimization (cf.~Eq.~\refeq{eq variation of F'}) yield the addition conditions of $\mathcal{\tilde{H}}\equiv 0$ for $\Delta s \vert_{poles} \neq 0$, where
\begin{equation}
    \label{eq Hamiltonian h0}
    \begin{split}
     \mathcal{\tilde{H}}&=\frac{r}{2}\left[\psi^2_s-\left(\frac{\sin \psi}{r}\right)^2\right]-\frac{\Sigma}{\kappa}r-\frac{1}{2}\frac{P}{\kappa} r^2 \sin \psi \\
     & +\gamma \cos \psi - \frac{F}{2 \pi \kappa} \sin \psi \,,
     \end{split}
\end{equation}
by definition $\mathcal{\tilde H}\equiv \mathcal{\tilde L}-\psi_s \partial \mathcal{\tilde L}/\partial \psi_s-r_s \partial \mathcal{\tilde L}/\partial r_s$, and the condition of zero moment at the poles 
\begin{equation}
    \label{eq zero moment h0}
    \left(r \psi_s+\sin \psi\right)\vert_{poles}=0\,,
\end{equation}
for $\Delta \psi \vert_{poles}\neq 0$. The numerical solution of the modified system of Euler-Lagrange equations determines the pressure, tension, and axial force for specified values of volume, area, and height, respectively; the length $L$ and the tilt-angle at the poles are determined self-consistently such that $\mathcal{\tilde H}$ is constant and the moment is zero. 

\section{Note on the correspondence between shape equations}
The direct correspondence between the general shape equation \refeq{Shape Eq normal} and Eqs.~\refeq{eq JulicherSeifert 1}-\refeq{eq JulicherSeifert 3} is obtained by eliminating the Lagrange multiplier functions ($\gamma,\eta$) from Eqs.~\refeq{eq JulicherSeifert 1}-\refeq{eq JulicherSeifert 3} and \refeq{eq Hamiltonian} using $\mathcal{H}\equiv 0$.

The steps are as follows: (i) eliminate $\eta=\eta(\psi,\psi_s,r,\gamma)$ from Eq.~\refeq{eq Hamiltonian} setting $\mathcal{H}=0$; (ii) this expression is then used in Eq.~\refeq{eq JulicherSeifert 1} to yield a relation for $\gamma=\gamma(\psi,\psi_s,\psi_{ss},r)$; (iii) finally, $\gamma$ is eliminated from Eq.~\refeq{eq JulicherSeifert 2} by differentiation with respect to arclength. This procedure results in a third-order shape equation in the tilt angle as previously reported in the literature \cite{JulicherSeifert1994}, that can be recast in the form of  Eq.~\refeq{Shape Eq normal} using definitions \refeq{eq meridional curvature}-\refeq{eq azimuthal curvature} and relation \refeq{eq mean curvature}. 

This equivalence between the general form of the shape equation \refeq{Shape Eq normal} and the system of Euler-Lagrange equations for axisymmetric vesicle shapes was a controversial topic in the 90s and early 2000s \cite{jian1993shape,naito1993counterexample,ZhengLiu1993,JulicherSeifert1994,podgornik1995parametrization,blyth2004solution}. Ou-Yang and coworkers \cite{jian1993shape,naito1993counterexample} argued that the Euler-Lagrange shape equations when parametrized by the radial distance from the symmetry axis to a point on the surface \cite{deuling1976curvature,svetina1989membrane}, or by arclength \cite{SeifertBerndlLipowsky1991} led to different shape equations when compared to the general shape equation \refeq{Shape Eq normal} specialized to axisymmetric geometries. Zheng \& Liu \cite{ZhengLiu1993} showed that both shape equations (see Eqs.(2) and (3) in Ref.\cite{ZhengLiu1993}) are relatable by a simple formula where the Euler-Lagrange shape equation is cast as a first integral of the more general, higher-order shape equation. In fact, both equations yield the same results for closed vesicles with smooth profiles where the constant of integration in Eq.(5) of Ref.\cite{ZhengLiu1993} is set to zero. This constant of integration can be associated with the axial point force discussed above \cite{bozic1997theoretical,derenyi2002formation} and hence vesicles with smooth, analytical contours are freely suspended or unconstrained. In this limit, vesicle contours are independent of the choice of parametrization, the total length and the height of the vesicle are free to vary, and the  resulting vesicle profiles reduce to a special subset of minimum energy, stationary solutions to Eq.~\refeq{Shape Eq normal} \cite{bozic1997theoretical}. The complementary, higher-energy set of vesicle shapes obtained from Eq.~\refeq{Shape Eq normal} lose analyticity at the poles where discontinuities in higher order derivatives of space variables are predicted \cite{podgornik1995parametrization,derenyi2002formation}. These non-analytic stationary shapes are associated with vesicle configurations resulting from the action of axial point forces, or, equivalently from an additional geometric constraint of fixed vesicle height \cite{JulicherSeifert1994,podgornik1995parametrization,guven2018geometry}. 

Blyth \& Pozrikidis \cite{blyth2004solution} revisited this topic and pointed out inconsistencies in the derivation of the Euler-Lagrange shape equations \refeq{eq JulicherSeifert 1}-\refeq{eq JulicherSeifert 3} with $\eta=0$ when the ``Hamiltonian" function of this system is set to zero. The authors presented numerical solutions to the general shape equation  \refeq{Shape Eq normal} for axisymmetric shapes and enforced smoothness of the profile at the poles by setting $dc_1/ds=0$ as one of the boundary conditions. Note that this is equivalent to setting the axial point force to zero (cf. Eq.~\refeq{eq axial force H}) which yields a special subset of stationary solutions to Eq.~\refeq{Shape Eq normal}. Hence, the results shown in Figs. 3(a)-(c) of Blyth \& Pozrikidis are for freely suspended vesicles and are in agreement with the results obtained from the system of Euler-Lagrange shape equations reported in Refs.~\cite{SeifertBerndlLipowsky1991,JulicherSeifert1994} with $\eta=0$. 

Blyth \& Pozrikidis \cite{blyth2004solution} also computed axisymmetric shapes using a thin-shell formulation for isotropic tensions and isotropic stress resultants integrated across the membrane thickness. Their results indicate a broader set of stationary shapes that arise from the solution of stress balance shape equations that are dynamically-equivalent, not exactly equal to Eq.~\refeq{Shape Eq normal} or Eq.~3 in Ref.~\cite{blyth2004solution}. The reason for this difference in form of the shape equations is a consequence of the choice of the linear constitutive equation for the meridional, $M_m$, and azimuthal, $M_\phi$, bending moments acting on a patch of membrane. For instance, for $M_m=\kappa c_1$ and $M_{\phi}=\kappa c_2$, the shape equations derived from force-torque balance assuming isotropic lateral tensions and isotropic integrated stresses derived in Ref.~\cite{blyth2004solution} differ from the shape equation \refeq{Shape Eq normal}; however, as shown in the Appendix of Powers et al. \cite{PowersPRE2002}, the general shape equation is recovered if the bending moments are defined in terms of the mean curvature,  i.e., $M_m=M_\phi=\kappa (c_1+c_2)$. 

Some vesicle profiles reported in Ref.~\cite{blyth2004solution} are in qualitative agreement with the solutions shown in section \ref{sec:results} of this work; however, self-intersection of the shapes in multi-lobed branches or pinching dynamics at the poles of vesicles with elongated tips are not verified herein for vesicle shapes with two-fold symmetry. This suggests that the physical conditions in both works are different within a higher dimensional configurational space; moreover, all numerical solutions in Ref.~\cite{blyth2004solution} are for unconstrained vesicles. 

\section{Local analysis of the tilt angle near the poles} \label{sec local analysis}
In this Appendix we show a local analysis of the shape equations \refeq{eq JulicherSeifert 1}-\refeq{eq JulicherSeifert 3} near the pole (i.e., for $\vert s\vert\ll 1$) where we take $\psi \to 0$ and $r\to 0$. Since $r_s=1$ at $s=0$, it follows from Eq.\refeq{eq drds dzds tilt}(a)  that $r\sim s$ to leading order. In this limit, Eq.~\refeq{eq JulicherSeifert 2} reduces to 
\begin{equation}
    \label{eq Gamma asymptotic}
    \gamma_s \sim \frac{\Sigma}{\kappa} + \frac{1}{2}\left[\psi^2_s-\frac{\psi^2}{r^2}\right]\,.
\end{equation}
Since $\mathcal{H}\equiv 0$, we assume $r\left[\psi^2_s-\frac{\psi^2}{r^2}\right]\to 0$ as $s\to 0$ yielding a linear, local behavior for the Lagrange multiplier function,
\begin{equation}
    \label{eq linear behavior gamma}
    \gamma\sim \frac{\Sigma}{\kappa} s\,,
\end{equation}
implying that $\gamma(0)=0$. Inspection of Eq.~\refeq{eq JulicherSeifert 1} in the limit as $\vert s \vert \ll 1$, leads to
\begin{equation}
    \label{eq psi_ss asymptotic}
    \psi_{ss} \sim \left(\frac{\psi}{s^2}-\frac{\psi_s}{s}+\frac{\eta}{s} \right)+\frac{1}{2}\frac{P}{\kappa} s + \frac{\gamma \,\psi}{s}\,.
\end{equation}
Since both $\gamma$ and $\psi$ tend to zero as $s \to 0$, we neglect the term $\sim (\gamma \psi)/s$; note that the pressure also vanishes as $s \to 0$. Thus, to leading order, the tilt-angle is governed by 
\begin{equation}
    \label{eq ode for psi}
    s^2 \psi_{ss} +s \psi_s - \psi = -\eta s \,,
\end{equation}
that admits a homogeneous solution of the form,
\begin{equation}
    \label{eq psi local}
    \psi \sim a s + \frac{b}{s}\,,
\end{equation}
where we set $b=0$ since $\psi\to 0$ as $s \to 0$. A particular solution to Eq.~\refeq{eq ode for psi} is 
\begin{equation}
    \label{eq psi particular}
    \psi_{p}=-\frac{1}{2} \eta \,s \ln s \,,
\end{equation}
and hence the general solution local to $s=0$ is 
\begin{equation}
    \label{eq tilt sol local s=0}
    \psi\sim a \, s -\frac{1}{2} \eta \,s \ln s \,.
\end{equation}
An equivalent local form for the tilt angle $\psi$ has been previously reported in Ref. \cite{bozic1997theoretical}. Inserting the local behavior for the tilt angle \refeq{eq tilt sol local s=0} into Eq.~\refeq{eq ode for psi} confirms, after integration, the leading order behavior of $\gamma$ given by Eq.~\refeq{eq linear behavior gamma} using $\gamma(0)=0$. In fact,
\begin{equation}
   \label{eq gamma local}
    \gamma \sim \gamma_0\, s +\gamma_1 \,s \ln s 
\end{equation}
where 
\begin{equation}
    \label{eq gamma_0}
    \gamma_0=\frac{\Sigma}{\kappa}-\frac{\eta}{2}\left(a+\frac{\eta}{4}\right)\,,
\end{equation}
and
\begin{equation}
    \label{eq gamma_1}
    \gamma_1=\frac{\eta^2}{4}\,.
\end{equation}

The local behavior for the spatial variables $(r,z)$ can be obtained directly from the local behavior of the tilt angle $\psi$. Inserting Eq.~\refeq{eq tilt sol local s=0} into relations \refeq{eq drds dzds tilt}, one gets after integration 
\begin{equation}
    \label{eq local behvavior r}
    r \sim  s +r_1 \,s^3 (\ln s)^2+r_2\, s^3 \ln s+r_3 \,s^3 + O(s^5 (\ln s)^4) \,,
\end{equation}
and 
\begin{equation}
    \label{eq local behvavior z}
    z \sim h_0 + z_1 \,s^2 \log s + z_2 \, s^2 + O(s^4 (\ln s)^3) \,,
\end{equation}
 where the $O(1)$ constant of integration in Eq.~\refeq{eq local behvavior r} is set to zero for closed shapes, $h_0$ is the height of the vesicle at $s=0$ (north pole) and the south pole is located at the origin of the coordinate system illustrated in Fig.~\ref{fig:coordinate system}(b) (i.e., $z(s_2)=0$). The asymptotic coefficients in Eqs.~\refeq{eq local behvavior r} and \refeq{eq local behvavior z} are  
 \begin{equation}
     \label{eq local coeffs r}
     \begin{split}
     r_1&=-\frac{1}{24}\eta^2\,, \quad r_2=\eta \left(\frac{a}{6}+\frac{\eta}{36}\right)\,, \\
     & \quad r_3=-\frac{a}{6} -\eta \left(\frac{a}{18}+\frac{\eta}{108}\right)\,,
     \end{split}
 \end{equation}
and 
\begin{equation}
    \label{eq local z1}
     z_1=\frac{\eta}{4}\,,\quad  z_2=-\left(\frac{a}{2}+\frac{\eta}{8}\right)\,,
\end{equation}
respectively.

Equations \refeq{eq local behvavior r}-\refeq{eq local behvavior z} show non-analytic behavior for the spatial variables $(r,z)$ near the poles. If $\eta=0$, $r$ and $z$ can be expressed as Taylor series expansions of cosine and sine about $s=0$, respectively, since $\psi \sim a s \to 0$ as $s \to 0$ and the logarithmic dependance is removed. In this case, the contours are considered smooth for all $s$.

A direct relation between the Lagrange multiplier $\eta$ and the axial force, $F$, is obtained using definition \refeq{eq axial force H} and the asymptotic behavior of the tilt angle \refeq{eq tilt sol local s=0}, yielding
\begin{equation}
    \label{eq relation eta and force}
    F=2 \pi \kappa \eta\,,
\end{equation}
where $H_s\sim \psi_{ss}$ and $r\sim s$. Smooth vesicle contours with local analytic behavior for $\vert s \vert\ll 1$ implies that the axial force vanishes at the poles (i.e., the vesicle is freely suspended) if, and only if, $H_s=0$.  Alternatively, if the Lagrange multiplier $\eta$ is nonzero and finite, the point-force acting at the poles is also nonzero and finite since $(H_s r)\vert_{poles}$ is bounded for shapes with finite energy. The non-analiticity of axisymmetric, closed contours when $\eta \neq 0$ has been pointed out in the literature in Refs. \cite{JulicherSeifert1994,podgornik1995parametrization}. Note that the local behavior of $\psi$ leading to $\gamma(0)=0$ implies that the ``Hamiltonian" of the system is constant and equal to zero for all $s$ even when $\eta$ is nonzero and finite. In this case, the axial force is sufficient to guarantee the interfacial force balance \refeq{Shape Eq normal} at the poles or, equivalently, to satisfy the condition that the first variation of the total elastic energy \refeq{eq Bending Energy point force} is zero for all $s$ (cf. Eq.~\ref{eq: force relation} in Appendix \ref{Appendix shape equations}).

\subsection{Effect of spontaneous curvature} \label{sec: local analysis C0}
The local analysis presented in Appendix \ref{sec local analysis} can be extended to include the effect of spontaneous curvature, where the shape equations \refeq{eq JulicherSeifert 1} and \refeq{eq JulicherSeifert 3} remain the same, and Eq.~\refeq{eq JulicherSeifert 2} becomes \cite{JulicherSeifert1994}
\begin{equation}
    \label{eq JulicherSeifert 2 C0}
     \gamma_{s}=\frac{1}{2}(\psi_s - C_0)^2 -\frac{\sin^2 \psi}{2\,r^2}+\frac{\Sigma}{\kappa}+\frac{P}{\kappa} r \sin \psi \,.
\end{equation}
Inserting the rescaled forms of the tilt angle and tension  
\begin{equation}
    \label{eq psi C0}
    \tilde \psi = \psi -C_0 r\,,
\end{equation}
and
\begin{equation}
    \label{eq psi C0 2}
    \tilde \Sigma = \Sigma -\frac{1}{2}\kappa C^2_0 \,,
\end{equation}
in Eq.~\refeq{eq JulicherSeifert 2 C0} yields
\begin{equation}
    \label{eq Gamma asymptotic C0}
    \gamma_s \sim \frac{\tilde \Sigma}{\kappa} - \frac{\tilde \psi}{r} C_0 +\frac{1}{2}\left[\tilde \psi^2_s-\frac{\tilde \psi^2}{r^2}\right]\,.
\end{equation}
We follow assumption \refeq{eq linear behavior gamma} where
\begin{equation}
    \label{eq linear behavior gamma C0}
    \gamma\sim \frac{\tilde \Sigma}{\kappa} s\,,
\end{equation}
is obtained by inspection of the Hamiltonian Eq.~\refeq{eq Hamiltonian} using the condition $\mathcal{H}\equiv 0$ given that 
\begin{equation}
    \label{eq condition H=0 1}
    r\left[\tilde \psi^2_s-\frac{\tilde \psi^2}{r^2}\right]\to 0 \,,
\end{equation}

\begin{equation}
    \label{eq condition H=0 2}
    \gamma \sim r \tilde \psi_s C_0 \,,
\end{equation}
and $\gamma(0)=0$. Hence, the governing equation for the rescaled tilt angle $\tilde \psi$ has the same form as in Eq.~\refeq{eq ode for psi} with solution given by 
\begin{equation}
    \label{eq tilt sol local s=0 C0}
    \tilde \psi \sim a \, s -\frac{1}{2} \eta \,s \ln s \,,
\end{equation}
where assumptions \refeq{eq condition H=0 1}-\refeq{eq condition H=0 2} are automatically satisfied. Substituting the rescaled solution for $\tilde \psi$ into Eq.~\refeq{eq Gamma asymptotic C0} yields the local behavior
\begin{equation}
   \label{eq gamma local C0}
    \gamma \sim \tilde \gamma_0\, s + \gamma_1 \,s \ln s + O(s^2 \ln s)
\end{equation}
where 
\begin{equation}
    \label{eq gamma_0 C0}
    \tilde \gamma_0=\frac{\tilde \Sigma}{\kappa}-\frac{\eta}{2}\left(a+\frac{\eta}{4}\right)\,,
\end{equation}
and $\gamma_1$ is given by Eq.~\refeq{eq gamma_1}.

\section{Experimental methods} \label{Appendix: Experimental Method}

All imaging is done using a high speed camera at 60 fps (Photron SA1, USA) and optical microscope (phase contrast Zeiss A1, Germany).

\subsection{Electroformation}
Giant unilamellar vesicles (GUVs) were formed from lipid, dioleoylphosphatidylcholine (DOPC). The lipids were purchased from Avanti Polar Lipids (Alabaster, AL) and Polymer Source Inc. (Montreal, Canada), respectively. Sucrose and Glucose were obtained from Sigma Aldrich, USA. HPLC water (22934 grade) was purchased from Fisher Scientific, USA. A small volume, 10 $\mu$l, of the
4 mM lipid solution concentration was spread on the conductive surface of two glass
slides coated with indium tin oxide (ITO) (Delta Technologies).
The glass slides were then stored under a vacuum for 1--2 hours
to remove traces of organic solvent. Afterwards, a 2 mm Teflon
spacer was sandwiched between the glass slides and the
chamber was gently filled with 40 mM sucrose solution.
The slides (conductive side facing inward) were connected to an
AC signal generator Agilent 33220A (Agilent Technology GmbH,
Germany). An AC field of voltage 1.5 V and frequency 10 Hz
applied for 2 hours at room temperature, resulting in 10-50 $\mu$m
sized vesicles. The harvested vesicles were diluted 10 times in 44 mM glucose solution to obtain fluctuating vesicles.

\subsection{Bending rigidity measurement} 
Membrane tension was probed using the flickering spectroscopy. The method takes advantage of non-invasive data collection and well-developed statistical analysis criteria. The details of the method are highlighted in Gracia et al. \cite{gracia2010effect} and Faizi et al. \cite{faizi2019bending}.  A time series of fluctuating vesicles at the equatorial cross section was recorded.  The fluctuating contour is represented in Fourier modes, $r(\phi)=a\left(1+\sum_q u_q(t) \exp(i q \phi)\right)$. The amplitude of the fluctuations $u_q$ 
can be presented with mean square amplitude that depends on the membrane bending rigidity $\kappa$ and tension $\Sigma$, $\langle \left|u_{q}\right |^2\rangle\sim\frac{k_bT}{\kappa \left(q^3+\bar \Sigma q\right)}$,
where $\kT$ is the thermal energy (k$_B$ is the Boltzmann constant and T is the temperature), $\bar\Sigma=\Sigma R^2/\kappa$, and $a$ is the initial radius of the vesicle. The integration time effect of the camera was minimized by acquiring images at a low shutter speed of 200 $\mu$s. At least 10,000 images were obtained for each vesicle for robust statistics.

\subsection{Electrodeformation}
The electrodeformation experiments are conducted in an electrofusion chamber (Eppendorf, Germany). The chamber is made from Teflon with two 92 $\mu$m cylindrical platinum parallel electrodes 500 $\mu$m apart. The field is applied using a function generator (Agilent 3320A, USA). The function generator is controlled using a custom built MATLAB (Mathworks, USA) program. This gives a precise control over the strength and duration of applied electric fields.

The image acquisition rate for electrodeformation recordings is kept to a constant of 60 fps for lipid vesicles and the shutter speed is 
fixed to 300 $\mu$ s. The time evolution of the vesicle is analyzed using a home-made image analysis software. The software uses a Fourier series to fit the vesicle contour, $r_s=\sum_{n=0}^\infty c_n\cos(n \theta)+d_n\sin(n \theta)$, where $r_{s}$ is the vesicle contour radius at the azimuthal angle $\theta$, $c_n$ and $d_n$ are the amplitude of the mode number $n$. The second mode in the series is used to determine the major and minor axis, $a_1$ and $b_1$, of the deformed vesicles to evaluate the aspect ratio $\alpha={(1+c_2)}/{(1-c_2)}$.

\subsection{Detailed balance analysis}
\label{Appendix: Detailed Balance}
\begin{figure*}
\centering
\includegraphics[scale=.48]{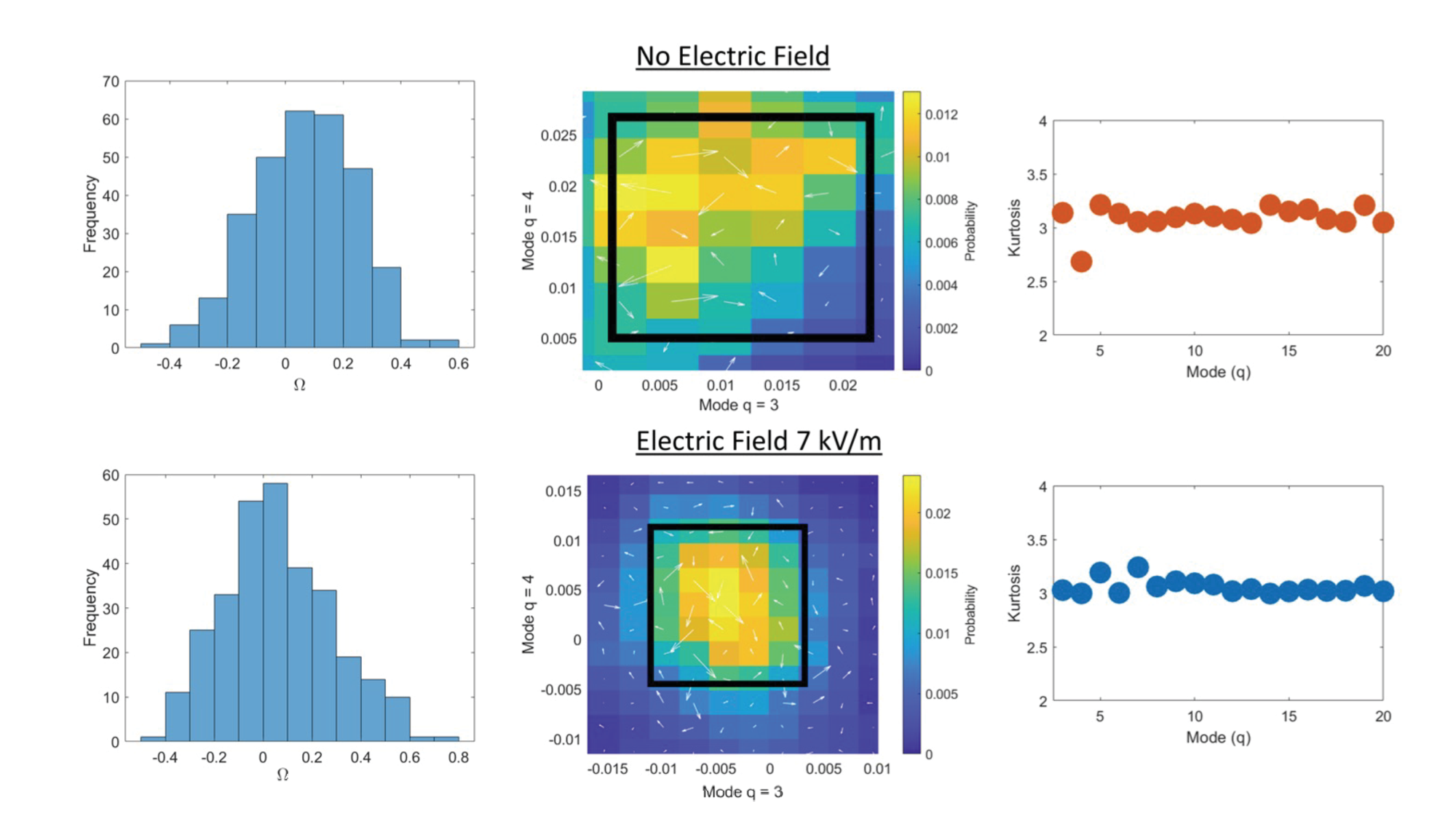} 
\vspace{-.55cm}
\caption{Nature of membrane fluctuations probed in the presence and absence of electric field. Probability current flux, $\Omega$, detailed balance, and Kurtosis values for DOPC vesicle in the absence of applied field strength (a), and in the presence of electric field at 7 kV/m, (b). The salt concentration in the inner and outer solutions for DOPC vesicles are 0.4 mM NaCl and 0.8 mM NaCl, respectively.}
\label{DB}
\end{figure*}

To check for the equilibrium nature of the fluctuations, we tested for broken detailed balance in the transitions between microscopic configurations based on height-height membrane fluctuations \cite{battle2016broken} (see chapter 6 of Ref.~\cite{Hammad2022} for more details about the method).
The configurations correspond to the shapes defined by different Fourier modes. In equilibrium, it is equally likely for the forward and backward transitions to occur between any two different Fourier modes. A non-equilibrium system, however, would display a probability flux in the phase space of shapes.  Figure \ref{DB} shows the probability density map for Fourier modes 3 and 4 of vesicles fluctuations in the absence and presence of electric field strength, as indicated. The probability is defined as the ratio of the time spent at a given state. The arrows indicate the currents across box boundaries determined by counting transitions between boxes. A nonzero value of the  contour integral of the probability current, $\Omega = \frac{\oint_{C} \textbf{j} \cdot d\textbf{l}}{\oint_{C} |\textbf{j}| \ dl}$, would indicate out of equilibrium dynamics. However, we noticed for moderate electric field strength ranging from 0-10 kV/m the detailed balance was not broken as indicated by $\Omega\sim\,0$. This implies that the fluctuations are still thermally driven in the presence of electric field as well.

We characterized the Gaussianity of  the fluctuations using the fourth PDF moment, Kurtosis, $\rm Kurt$. For a Gaussian distribution, $\rm Kurt=3$. In Figure \ref{DB} we demonstrated the Kurtosis for every mode number for the same vesicle in presence (7 kV/m) and absence of electric field strength. Our results confirm the previous analysis of unbroken Detailed balance with Kurtosis values $\rm Kurt \sim3$ for membrane fluctuations in the presence of electric field as well. 

\bibliography{apssamp}

\end{document}